\documentclass{emulateapj}

\usepackage{amsmath}
\usepackage[]{natbib}
\usepackage[usenames]{color} 

\newcommand{\kpc}{\>{\rm kpc}}
\newcommand{\kms}{\>{\rm km}\,{\rm s}^{-1}}

\newcommand{\pc}{\>{\rm pc}}
\newcommand{\sig}[1]{\mbox{$\sigma_{#1}$}}
\newcommand{\Msun}{\mbox{$\rm M_{\odot}$}}
\newcommand{\Mbh}{\mbox{$\rm M_{\bullet}$}}
\newcommand{\Mbul}{\mbox{$\rm M_{bul}$}}
\newcommand{\Mgal}{\mbox{$\rm M_{gal}$}}

\newcommand{\Msig}{\mbox{$\rm M_{\bullet}-\sig{e}$}}
\newcommand{\MBHsig}{\mbox{$\rm M_{\rm BH}-\sig{e}$}}
\newcommand{\Mmb}{\mbox{$\rm M_{\bullet}-M_{bul}$}}
\newcommand{\Mlb}{\mbox{$\rm M_{\bullet}-L_{bul}$}}
\newcommand{\Mnb}{\mbox{$\rm M_{\bullet}-n$}}

\newcommand{\dMbh}{\mbox{$\Gamma_{\bullet}$}}

\newcommand\degrees{^\circ}
\newcommand{\db}{\mbox{$\rm D/B$}}
\newcommand{\bd}{\mbox{$\rm B/D$}}
\newcommand{\rd}{\mbox{$R_{\rm d}$}}
\newcommand{\zd}{\mbox{$z_{\rm d}$}}
\newcommand{\re}{\mbox{$R_{\rm e}$}}
\newcommand{\rei}{\mbox{$R_{\rm e,0}$}}
\newcommand{\rdrei}{\mbox{$R_{\rm d}/R_{\rm e,0}$}}
\newcommand{\rdre}{\mbox{$R_{\rm d}/R_{\rm e}$}}
\newcommand{\rerd}{\mbox{$R_{\rm e}/R_{\rm d}$}}

\newcommand{\etal}{{et al.~}}

\newcommand\ie{{\it i.e.}}

\newcommand{\cut}[1]{}

\slugcomment{{\it Unfinished draft} \today}

\shorttitle{Disk Assembly and the $\rm M_{\rm BH}-\sig{e}$ Relation
of Supermassive Black Holes}

\shortauthors{Debattista \etal}

\begin{document}

\title{Disk Assembly and the \MBHsig\ Relation of Supermassive Black
  Holes}

\author{
Victor P. Debattista\altaffilmark{1,2}, 
Stelios Kazantzidis\altaffilmark{3},
Frank C. van den Bosch\altaffilmark{4,5}
}

\altaffiltext{1}{RCUK Fellow, Jeremiah Horrocks Institute, University
of Central Lancashire, Preston, PR1 2HE, UK {\tt vpdebattista@gmail.com}}

\altaffiltext{2}{Visiting Lecturer, Department of Physics, University
of Malta, Tal-Qroqq Street, Msida, MSD 2080, Malta}

\altaffiltext{3}{Center for Cosmology and Astro-Particle Physics; and
Department of Physics; and Department of Astronomy, The Ohio State
University, Columbus, OH 43210, USA; {\tt stelios@mps.ohio-state.edu}}

\altaffiltext{4}{Astronomy Department, Yale University, PO Box 208101,
New Haven, CT 06520-8101, USA {\tt frank.vandenbosch@yale.edu}}

\altaffiltext{4}{Lady Davis Visiting Professor, Racah Institute of Physics, 
Hebrew University, Jerusalem 91904, Israel}

\begin{abstract} 
  Recent Hubble Space Telescope ({\it HST}) observations have revealed
  that a majority of active galactic nuclei (AGN) at $z \sim 1-3$ are
  resident in isolated disk galaxies, contrary to the usual
  expectation that AGN are triggered by mergers.  Here we develop a
  new test of the cosmic evolution of supermassive black holes (SMBHs)
  in disk galaxies by considering the local population of SMBHs.  We
  show that substantial SMBH growth in spiral galaxies is {\it
    required} as disks assemble.  SMBHs exhibit a tight relation
  between their mass and the velocity dispersion of the spheroid
  within which they reside, the \Msig\ relation.  In disk galaxies the
  bulge is the spheroid of interest.  We explore the evolution of the
  \Msig\ relation when bulges form together with SMBHs on the \Msig\
  relation and then slowly reform a disk around them.  The formation
  of the disk compresses the bulge raising its \sig{e}.  We present
  evidence for such compression in the form of larger velocity
  dispersion of classical bulges compared with elliptical galaxies at
  the same mass.  This compression leads to an offset in the \Msig\
  relation if it is not accompanied by an increased \Mbh.  We quantify
  the expected offset based on photometric data and show that, on
  average, SMBHs must grow by $\sim 50-65\%$ just to remain on the
  \Msig\ relation.  We find no significant offset in the \Msig\
  relations of classical bulges and of ellipticals, implying that
  SMBHs have been growing along with disks.  Our simulations
  demonstrate that SMBH growth is {\it necessary} for the local
  population of disk galaxies to have remained on the \Msig\ relation.
\end{abstract}

\keywords{black hole physics --- galaxies: bulges --- galaxies:
  evolution --- galaxies: nuclei --- cosmology: theory --- methods:
  numerical}

%%%%%%%%%%%%%%%%%%%%%%%%%%%%%%%%%%%%%%%%%%%%%%%%%%%%%%%%%%%%%%%%%%%%%

\section{Introduction}
\label{sec:intro}

The energetics and demographics of active galactic nuclei (AGN), which
are found already by $z = 6$ \citep{zhe_etal_00, fan_etal_00,
  fan_etal_04, fan_etal_06}, can be explained by the presence of
accreting supermassive black holes \citep{lynden_69, soltan_82,
  cho_tur_92, sal_etal_99, mer_fer_01b}.  The {\it Hubble Space
  Telescope} ({\it HST}) has revealed such supermassive black holes
(SMBHs), with masses in the range $10^6 - 10^9 \Msun$, in a number of
nearby quiescent galaxies \citep{kor_ric_95}.  However the formation
and growth of SMBHs remains something of a mystery with a variety of
models proposed \citep{ mad_ree_01, por_mcm_02, mil_ham_02,
  oh_haiman02, vol_etal_03, isl_etal_03, kou_etal_04, hop_etal_06,
  volonteri_rees05, beg_etal_06, lod_nat_06, kin_pri_06, mayer+10}.

SMBHs exhibit a number of scaling relations, the tightest of which is
the \Msig\ relation between their mass, \Mbh, and the velocity
dispersion, \sig{e}, of the spheroids within which they reside.  A
scaling-relation of the form $\log\Mbh = \alpha +
\beta\log(\sig{e}/200 \kms)$ was found by \citet{geb_etal_00} and
\citet{fer_mer_00}.  Early measurements of the slope $\beta$ varied
from $4.02 \pm 0.32$ \citep{tre_etal_02} to $4.72 \pm 0.36$
\citep{mer_fer_01}.  More recent measurements still find a large range
of $\beta$ spanning $\beta=4.24\pm0.41$ \citep{gultekin+09} to $\beta
= 5.57 \pm 0.33$ \citep{mcconnell+11, mcconnell_ma12}.  Other
suggested correlations between SMBHs and their host bulges include the
\Mlb\ or \Mmb\ relations with the bulge luminosity or mass
\citep{kor_ric_95, mag_etal_98, mar_hun_03, hae_rix_04}, and the \Mnb\
relation with the bulge S\'ersic index \citep{gra_dri_07}.  A
three-parameter fundamental plane for SMBHs has also been suggested
\citep{mar_hun_03, def_etal_06, all_ric_07, hop_etal_07obs,
  hop_etal_07the, bar_kem_07}.  \citet{graham_08} argued that the
fundamental plane is caused by barred galaxies.  \citet{beifiori+12}
found that the fundamental plane is strongly dominated by \sig{e}.  A
relation between \Mbh\ and \Mgal, the mass of the host galaxy, has
been suggested \citep{ferrar_02, bae_etal_03, pizzella+05}.  Early
work found that the same relation is satisfied also by nuclear star
clusters \citep{fer_etal_06, weh_har_06, rossa+06}, which may provide
a unified picture of the growth of central massive objects
\citep{mcl_etal_06, markus+11}.  However more recent work has found
that nuclear star clusters and SMBHs follow different scaling
relations \citep{erwin_gadotti12, leigh+12, scott_graham12}.  This may
possibly be a result of the scaling relations being different in
late-type galaxies \citep{greene+10, erwin+gadotti12}.

The clues to SMBH growth and formation implied by these scaling
relations are non-trivial to decipher: The sphere of influence of a
typical SMBH has a radius of a few parsecs, which is some 2-3 orders
of magnitude smaller than the effective radius of a typical bulge.
What mechanism then gives rise to these scaling relations?  Do SMBHs
regulate bulge growth or is the growth of SMBHs restricted by the
bulge in which they reside?  Examples of the latter are scenarios in
which gas accretion onto the SMBH is regulated by star formation
\citep{bur_sil_01, kaz_etal_05, zheng+09}, or by stellar feeding of
SMBH accretion disks \citep{mir_kol_05}.  AGN feedback via heating,
pressure-driven winds or ionization typically gives rise to scenarios
in which SMBHS regulate their own or bulge growth \citep{sil_ree_98,
  wyithe_loeb_03, aking_03, mur_etal_05, dim_etal_05, saz_etal_05,
  you_etal_08}.  High velocity outflows that may be associated with
such AGN feedback have been observed in Seyfert 1 galaxies
\citep[e.g.][]{crenshaw+99}. In addition, in semi-analytic models, AGN
feedback is often invoked to explain the high mass end of the
luminosity function \citep{cro_etal_06, bow_etal_06}.  Alternatively,
collapse models in which the \Msig\ relation is an indirect
consequence of unrelated processes have also been proposed
\citep{hae_kau_00, ada_etal_01, ada_etal_03, peng07,
  volonteri+natarajan09, jahnke_maccio11}.

One of the characteristics of the scaling relations is that \Mbh\
correlates with the properties of the host {\it spheroid}.  In disk
galaxies this is the bulge component.  Bulges in disk galaxies come in
two types: "classical" and "pseudo" bulges, with mixed types also
possible \citep[e.g.][]{erwin+03, deb_etal_05, athana_05, nowak+10}.
Classical bulges are believed to form via merging of sub-galactic
clumps, satellites and clusters \citep{els_62, tre_etal_75,
  sea_zin_78, kau_etal_93, bau_etal_96, vdbos_98}.  In essence
classical bulges are elliptical galaxies around which a disk has
reformed \citep[e.g.][]{steinmetz_navarro_02} although continued late
growth of classical bulges is also possible
\citep[e.g.][]{hopkins+10}.  Pseudo bulges instead are formed by the
secular evolution of disk structure, such as bars and spirals
\citep{com_san_81, com_etal_90, rah_etal_91, nor_etal_96, cou_etal_96,
  bur_fre_99, deb_etal_04, athana_05, dro_fis_07}.  \citet{kor_ken_04}
reviewed the observational evidence for pseudo bulge formation.  In
contrast to pseudo bulges, classical bulges form early, predating the
formation of the disk.  The difference between classical and pseudo
bulges is reflected also in their SMBH demographics.
\citet{gadotti_kauffmann_09} estimate that classical bulges account
for $41\%$ of the black hole mass in the local universe, while pseudo
bulges host only 4\%.  \citet{gultekin+09} find that the classical
bulges (including elliptical galaxies) follow the same \Msig\ relation
as the general population, with a scatter of $0.45 \pm 0.066$.
Instead \citet{hu_08} found that pseudo bulges have an \Msig\ relation
with the same slope as, but lower zero-point than, classical bulges.
\citet{greene+10} showed that the SMBHs of late-type galaxies, which
predominantly contain pseudo bulges, scatter below the \Msig\
relation.

A number of studies using simulations have explored the evolution of
the \Msig\ relation during hierarchical merging
\citep[e.g.][]{kaz_etal_05, spr_etal_05, spr_etal_05b, you_etal_08,
  joh_etal_09, rob_etal_06}.
However, recent {\it HST} observations have shown that a surprisingly
substantial fraction of AGN activity at high redshifts is associated
with isolated disk galaxies, rather than with mergers.
\citet{schawinski+11} show that $\sim 80\%$ of X-ray-selected AGN at
$z = 1.5-3$ are in low S\'ersic-index galaxies, indicative of disks.
They find that moderate luminosity AGN hosts at $z \sim 2$ are similar
to those at $z \sim 0$.  Excluding the high luminosity quasars, which
are triggered by mergers \citep{treister+10}, they estimate that
$23-40\%$ of SMBH growth occurs in intermediate brightness Seyfert
AGN.  The X-ray-selected sample of moderate-luminosity AGN at $1.5 < z
< 2.5$ of \citet{cisternas+11} consists of more than $50\%$ disk
galaxies, with ongoing mergers evident no more frequently than in
non-active galaxies.  \citet{schawinski+12} show that even heavily
obscured quasars are hosted largely by disks, not by mergers.  Studies
of star-formation using {\it Herschel} find that the specific star
formation rates of X-ray selected AGN hosts are no different from
those of inactive galaxies, also indicating that AGN hosts are not
undergoing fundamentally different behaviors \citep{mullaney+12a,
  mullaney+12b}.  Using multiwavelength surveys of AGN across
redshifts $0\le z \le 3$ \citet{treister+12} found that only the most
luminous AGN phases are connected to major mergers, the rest being
driven by secular processes.  The merger driven AGN activity accounts
for only $\sim 10\%$ of AGN.  The ``anti-hierarchical'' nature of
galaxy and AGN growth --- both the largest galaxies \citep{bower+92,
  thomas+05, nelan+05} and the brightest AGN \citep{ueda+03,
  hasinger+05} form at high redshift whereas lower mass galaxies and
moderate luminosity AGN peak at lower redshifts --- also hints that
internal evolution rather than mergers is the main driver of SMBH
growth.  AGN activity continues to be dominated by disk galaxies down
to the present: since $z\sim 1$ more than $85\%$ of AGN activity is
hosted in galaxies with no evidence of recent mergers
\citep{kocevski+12}.  Lastly, the presence of AGN in bulgeless
galaxies, which are thought to not have experienced much hierarchical
merging, provides further evidence that internal evolution is capable
of driving SMBH growth \citep{simmons+12, arayasalvo+12}.

This paper introduces a novel approach to exploring the origin of the
\Msig\ relation.  We study the consequences of disk regrowth for the
\Msig\ relation of classical bulges under the assumption that a
classical bulge forms with a SMBH satisfying the \Msig\ relation and
then a disk reassembles around it.  Growth of the disk then compresses
the bulge \citep{andred_98}.  Since \sig{e}\ is not an adiabatic
invariant, compression leads to its evolution, which we quantify here.
We study the effect of this evolution on the \Msig\ relation.  If
SMBHs remain on the \Msig\ relation then this implies that SMBH growth
is governed by the potential (as characterized by \sig{e}) within
which they sit, which is most likely if AGN feedback regulates SMBH
growth.  If instead we find that bulges evolve away from the \Msig\
relation, with the SMBHs retaining a memory of the bulge within which
they formed, then this implies that bulge growth is limited by the
SMBH, as would happen if AGN feedback quenches star formation in the
bulge.  The paper is organized as follows.  Section \ref{sec:simul}
describes the simulation methods used in this paper.  Section
\ref{sec:dispersions} presents the evolution of \sig{e}\ caused by
disk (re-)assembly and derives a photometric estimate for the increase
in \sig{e}.  In Section \ref{sec:msig} we predict the consequences of
bulge compression for the \Msig\ relation.  We find that the main
effect is a shift to lower mass in the zero-point of the relation.
Then in Section \ref{sec:offsets} we test this prediction on
observational data.  We find no evidence for such a shift, indicating
that SMBHs have grown along with disks.  We show that the degree by
which SMBHs must have grown is consistent with the new {\it HST}
estimates.  Section \ref{sec:conclusions} sums up our results.

%%%%%%%%%%%%%%%%%%%%%%%%%%%%%%%%%%%%%%%%%%%%%%%%%%%%%%%%%%%%%%%%%%%%%

\section{Numerical Simulations}
\label{sec:simul}

We construct initially spherically symmetric two-component galaxy
models consisting of a stellar bulge embedded in an extended dark
matter (DM) halo. For the DM component, we consider the cuspy,
cosmologically-motivated \citet[][hereafter NFW]{nfw96} density
profile given by
\begin{equation}
   \rho_{\rm DM}(r) = \frac{\rho_s} {\left (r/r_s\right) 
     \left (1 + r/r_s\right )^2} \qquad\hbox{($r \leq r_{\rm vir}$)},
   \label{density}
\end{equation}
where $\rho_s$ is a characteristic inner density, $r_s$ denotes the
scale radius of the density profile defined as the distance from the
center where the logarithmic slope, $\rm d \ln \rho(r)/\rm d \ln r$,
is equal to $-2$, and $r_{\rm vir}$ is the virial radius defined as
the radius enclosing an average density equal to the virial
overdensity times the critical density for a flat universe.  We adopt
the $\Lambda$CDM concordance cosmology with $\Omega_{\rm m}=0.3$,
$\Omega_{\Lambda}=0.7$, and $h=0.7$, and assume $z=0$.  The virial
overdensity is then equal to $\Delta_{\rm vir} \simeq 103.5$
\citep[e.g.,][]{lacey_cole93}.

The NFW density profile is formally infinite in extent with a
cumulative mass that diverges as $r \rightarrow \infty$. In order to
keep the total mass finite, we implement an exponential cutoff which
sets in at the virial radius and turns off the profile on a scale
$r_{\rm decay}$. The truncation scale $r_{\rm decay}$ is a free
parameter and controls the sharpness of the transition. Explicitly, we
model the density profile beyond $r_{\rm vir}$ by
\begin{equation}
   \rho(r)=\frac{\rho_{\rm s}} {c (1+c)^2} 
   \left(\frac{r}{r_{\rm vir}}\right)^{\kappa}
   \exp\left[-\frac{r-r_{\rm vir}}{r_{\rm decay}}\right]  
   (r>r_{\rm vir}),
   \label{exp_cutoff}
\end{equation}
where $c \equiv r_{\rm vir}/r_{\rm s}$ is the concentration parameter
and $\kappa$ is fixed by the requirement that $\rm d \ln \rho(r)/\rm d
\ln r$ is continuous at $r_{\rm vir}$. This procedure is necessary
because sharp truncations result in models that are not in equilibrium
\citep{kmm04}. For the purposes of the present study, we adopt a
concentration parameter $c=10$, appropriate for Milky Way Galaxy-sized
dark matter halos \citep{bullock+01}, and a truncation scale $r_{\rm
decay} = 0.1 r_{\rm s}$.

For the spatial distribution of the bulge component, we adopt the
de-projected S\'ersic law \citep{sersic_68} of \citet{sim_pra_04}:
\begin{equation}
   \rho(s) = \rho_0 \int_{0}^{1} \frac{\exp \left[-k s^\frac{1}{n} (1-x^{2})^{-\frac{1}{n-1}}
   \right]}{1-(1-x^{2})^{\frac{n}{n-1}}} \ x \ {\rm d}x \qquad\hbox{($n>1$)},
\end{equation}
where
\begin{equation}
   \rho_0 = \frac{k}{\pi} \frac{\Sigma_0}{R_{\rm e,0}} \frac{2}{n-1} \frac{1}{s^{\frac{n-1}{n}}}.
\end{equation}
In the above equations, $n$ denotes the S\'ersic index, $R_{\rm e,0}$
is the effective radius, \ie\ the radius that encloses half the total
projected luminosity, $\Sigma_0$ is the central value of the projected
mass profile, and $s \equiv r/R_{\rm e,0}$.  For $n \geq 1$, $k$ can
be estimated (with an error smaller than $0.1\%$) by the relation $k=
2n-0.324$ \citep{ciotti91}.
Our initial bulge has S\'ersic index $n=4$, \ie\ it is characterized
by a \citet{devauc_48} profile. For the specific galaxy model we
consider, the ratio between the mass of the bulge and the {\it virial}
mass of the halo is equal to $8 \times 10^{-3}$, while the ratio
between the bulge effective radius and the halo scale radius, $\rei/r_s
= 0.02$.  Our standard scaling has $\rei = 500$ pc and bulge mass
$M_b = 8 \times 10^9~ \Msun$, leading to $r_{\rm vir} = 270$ kpc and
$M_{\rm vir} = 10^{12} \Msun$.  Because we do not consider
non-gravitational processes such as gaseous dissipation, the
scale-free nature of gravity allows the rescaling of our models.

Monte-Carlo realizations of the $N$-body galaxy model are constructed
according to the procedure described in \citet{kmm04}, which is based
on sampling the exact phase-space distribution function (DF).  Under
the assumption of isotropy, the DF of each component depends only on
the binding energy per unit mass $E$:
\begin{equation}
   f_{i}(E) =\frac{1}{\sqrt{8} \pi^2} \left[\int_{0}^{E}
   \frac{{\rm d}^2 \rho_{i}}{{\rm d} \Psi^2} \frac{{\rm d} \Psi}{\sqrt{E-\Psi}} + 
   \frac{1}{\sqrt{E}} \left (\frac{{\rm d} \rho_{i}} {{\rm d}\Psi}\right)_{\Psi=0}\right],
   \label{two_comp.DF}
\end{equation}
where $\rho_{i}$ is the density profile of component $i$ and $\Psi(r)
= \psi_{\rm DM}(r)+\psi_{\rm stars}(r)$ is the total {\it relative}
gravitational potential. Note that the second term on the right-hand
side in Eqn. \ref{two_comp.DF} vanishes for any sensible behavior of
$\Psi(r)$ and $\rho_{i}(r)$ as $r \rightarrow \infty$.

The system generated this way needs to be softened; the gravitational
softening lengths are set to $\epsilon = 15$ pc for all particles
(including the disk particles described below) in all runs.  Since
softening the potential is equivalent to smoothing the density
distribution \citep{barnes12}, the initial conditions set up without
softening are not a perfect equilibrium. We therefore relax the
initial bulge$+$halo system for 250 Myr before we start growing the
disk.  During this period the bulge settles to a new equilibrium.  For
the remainder of this paper we refer to this relaxed model as the
initial conditions.

After the bulge$+$halo system has reached equilibrium, we investigate
its response to the growth of various external disk fields. The
growing disks follow an exponential distribution in cylindrical radius
$R$, and their structure is modeled as \citep{spitze_42, freema_70}:
\begin{equation}
  \rho_d(R,z,t) = \frac{m_d(t)}{8\pi z_d R_d^2} \exp\left(-\frac{R}{R_d}\right) {\rm
  sech}^2\left(\frac{z}{2 z_d}\right),
   \label{disc_density}
\end{equation}
where $m_d$, $R_d$, and $z_d$ denote the mass, radial scale-length,
and vertical scale-height of the disk, respectively.  Except for one
model, we use $z_d=0.15\,R_d$ in all experiments, a choice which is
consistent with observations of external galaxies
\citep{vanderkruit+searle82, degrijs+vanderkruit96}.  However, the
observed scatter in $z_d/R_d$ is quite substantial, reaching to values
as small as 0.05.  We show below that thinner disks lead to even
stronger compression.  Thus our assumption of $z_d=0.15\,R_d$ is
conservative.  We implicitly assume that the classical bulge is fully
formed at the last major merger.  \citet{hernquist_mihos95} showed
that minor mergers drive gas to small radii leading to a burst of star
formation and bulge growth.  However dissipation is now thought to
largely give rise to pseudo, not classical, bulges.  The origin of
bulges in high mass galaxies remains contentious.  \citet{weinzirl+09}
used bulge$+$disk$+$bar decompositions to argue that mergers cannot
account for the majority of bulges in current high mass galaxies.
\citet{hopkins+10} instead argued that major mergers dominate the
formation of these bulges, with minor mergers contributing another
$\sim 30\%$.  Nonetheless a separation between classical and pseudo
bulges seems to be well established, with observational evidence
indicating that properties such as morphologies, star formation rates
and correlations with disk properties, including color, change across
$n \simeq 2$ \citep{dro_fis_07, fis_dro_08, fisher+09,
  fisher_drory10}.

Each growing-disk simulation is performed by growing linearly over
time the mass of an initially massless Monte Carlo particle
realization of the desired disk model: $m_d(t) = (t/\tau) M_d$.  In
all simulations, we set $M_d = 1.6\times 10^{11} \Msun$ and $\tau = 2$
Gyr.  During the experiments the disks are held rigid with their
particles fixed in place, while both the bulge and halo particles are
live, allowing them to remain in equilibrium as the disk mass grows.
Throughout the experiments, all other properties of the growing disks
(e.g., scale-lengths, scale-heights) are kept constant.  Additional
details of this technique can be found in \citet{debattista+08},
\citet{villalobos+10} and \citet{kazantzidis+10}.  Our simulations do
not include a SMBH since the sphere of influence of the SMBH that
would correspond to the velocity dispersion of the initial bulge, $r_h
= G \Mbh/\sig{e}^2$ is only $\sim 4 \pc$, which is considerably
smaller than \rei.

The initial conditions of the two-component galaxy contain a total of
$4.4$ million particles ($4 \times 10^6$ dark matter particles and $4
\times 10^5$ bulge particles).  The disk is modeled with a further $4
\times 10^5$ particles.  Particles are set up using a quiet start
procedure \citep{sellwood83} that ensures that all components have
zero net momentum.  We set up particles in groups of four: the first
particle has $(x,y,z,v_x,v_y,v_z)$ while the rest have
$(-x,-y,z,-v_x,-v_y,v_z)$, $(-y,x,-z,-v_y,v_x,-v_z)$ and
$(y,-x,-z,v_y,-v_x,-v_z)$.

We run $5$ simulations, with varying ratio of disk scale-length to
initial bulge effective-radius: \rdrei\ = 0.5, 1, 2, 5, and 10.  We
save outputs at 25 Myr intervals corresponding to increments $\delta
M_d = 0.25 M_b$.  Table~\ref{tab:model_parameters} provides a summary
of a representative subset of the outputs.

All numerical simulations are carried out with the parallel $N$-body
code {\sc pkdgrav} \citep{stadel_phd}.  In all experiments, we set the
base timestep $\Delta t = 1.25$ Myr with timesteps refined such that
$\delta t = \Delta t/2^p < \eta (\epsilon/a)^{1/2}$, where $\epsilon$
is the softening and $a$ is the acceleration at a particle's current
position, with rung number $p$ as large as 29 allowed.  For all
simulations we set $\eta = 0.02$ and use an opening angle of the
treecode $\theta = 0.7$.  In the $\rdrei=1$ model, timesteps for
particles get as small as $2^{-10}$ of the base timestep, (\ie\ 1220
years).

\begin{table}[!ht]
\begin{centering}
\begin{tabular}{cc|ccc} \\ \hline
\multicolumn{1}{c}{\db} &
\multicolumn{1}{c|}{\rdrei} &
\multicolumn{1}{c}{\rdre} &
\multicolumn{1}{c}{$\sig{e}$} &
\multicolumn{1}{c}{$\dMbh$}  \\ 
    &    &  &         [$\kms$]    &   \\ \hline 
  0   &  - & -     & $ 115.3 \pm 0.3 $ &   -  \\
 0.25 &0.5 &  0.61 & $ 126.9 \pm 0.7 $ &  1.5 \\
 0.5  &0.5 &  0.70 & $ 137.7 \pm 1.2 $ &  2.0 \\
 1    &0.5 &  0.83 & $ 155.9 \pm 0.9 $ &  3.3 \\
 5    &0.5 &  1.33 & $ 229.4 \pm 2.3 $ & 15.7 \\
 10   &0.5 &  1.62 & $ 279.1 \pm 3.4 $ & 34.4 \\
 15   &0.5 &  1.82 & $ 316.7 \pm 4.1 $ & 56.9 \\
 20   &0.5 &  1.98 & $ 345.2 \pm 3.8 $ & 80.4 \\
 0.25 &  1 &  1.12 & $ 119.8 \pm 0.3 $ &  1.2 \\
 0.5  &  1 &  1.22 & $ 124.5 \pm 0.3 $ &  1.4 \\
 1    &  1 &  1.37 & $ 134.4 \pm 1.8 $ &  1.8 \\
 5    &  1 &  1.99 & $ 177.4 \pm 3.2 $ &  5.6 \\
 10   &  1 &  2.38 & $ 207.6 \pm 5.1 $ & 10.5 \\
 15   &  1 &  2.63 & $ 231.1 \pm 6.8 $ & 16.2 \\
 20   &  1 &  2.83 & $ 251.7 \pm 9.6 $ & 22.7 \\
 0.25 &  2 &  2.10 & $ 116.2 \pm 0.3 $ &  1.0 \\
 0.5  &  2 &  2.20 & $ 117.8 \pm 0.6 $ &  1.1 \\
 1    &  2 &  2.36 & $ 121.9 \pm 1.4 $ &  1.2 \\
 5    &  2 &  3.07 & $ 142.8 \pm 2.9 $ &  2.4 \\
 10   &  2 &  3.53 & $ 161.4 \pm 4.6 $ &  3.8 \\
 15   &  2 &  3.84 & $ 174.9 \pm 5.2 $ &  5.3 \\
 20   &  2 &  4.07 & $ 186.3 \pm 4.9 $ &  6.8 \\
 0.25 &  5 &  5.05 & $ 114.3 \pm 0.3 $ &  1.0 \\
 0.5  &  5 &  5.11 & $ 113.9 \pm 0.8 $ &  1.0 \\
 1    &  5 &  5.24 & $ 115.4 \pm 0.8 $ &  1.0 \\
 5    &  5 &  5.96 & $ 122.4 \pm 0.5 $ &  1.3 \\
 10   &  5 &  6.45 & $ 128.6 \pm 1.6 $ &  1.5 \\
 15   &  5 &  6.81 & $ 134.6 \pm 1.6 $ &  1.9 \\
 20   &  5 &  7.11 & $ 138.8 \pm 2.1 $ &  2.1 \\
 0.25 & 10 & 10.02 & $ 113.2 \pm 0.4 $ &  0.9 \\
 0.5  & 10 & 10.04 & $ 113.4 \pm 0.9 $ &  0.9 \\
 1    & 10 & 10.14 & $ 113.7 \pm 0.5 $ &  0.9 \\
 5    & 10 & 10.68 & $ 116.9 \pm 0.8 $ &  1.1 \\
 10   & 10 & 11.22 & $ 119.4 \pm 0.8 $ &  1.2 \\
 15   & 10 & 11.56 & $ 122.2 \pm 1.3 $ &  1.3 \\
 20   & 10 & 11.80 & $ 123.5 \pm 1.3 $ &  1.3 \\ \hline
\end{tabular}

\caption{
A representative sampling of the results at different times
(corresponding to different \db) of the simulations presented in this
paper.  The first two columns list the `input' parameters, where \db\
is the disk-to-bulge mass ratio and \rdrei\ is the ratio of disk
scale-length to initial bulge effective radius.  The last three
columns list the output parameters of the simulations.  \rdre\ is the
ratio of disk scale-length to final bulge effective radius, \sig{e}\
is the velocity dispersion of the bulge and \dMbh\ is the ratio of
final to initial \Mbh\ assuming the system starts and ends on the
  same \Msig\ relation with slope $\beta = 4$.  Both \rdre\ and
\sig{e}\ include the effects of relaxation of the initial conditions.
The first row corresponds to the initial conditions before the disk is
grown.
\label{tab:model_parameters}}
\end{centering} 
\end{table}

%%%%%%%%%%%%%%%%%%%%%%%%%%%%%%%%%%%%%%%%%%%%%%%%%%%%%%%%%%%%%%%%%%%%%

\section{Evolution of Velocity Dispersion}
\label{sec:dispersions}

\begin{figure}[!ht]
\centerline{
\includegraphics[angle=-90.,width=\hsize]{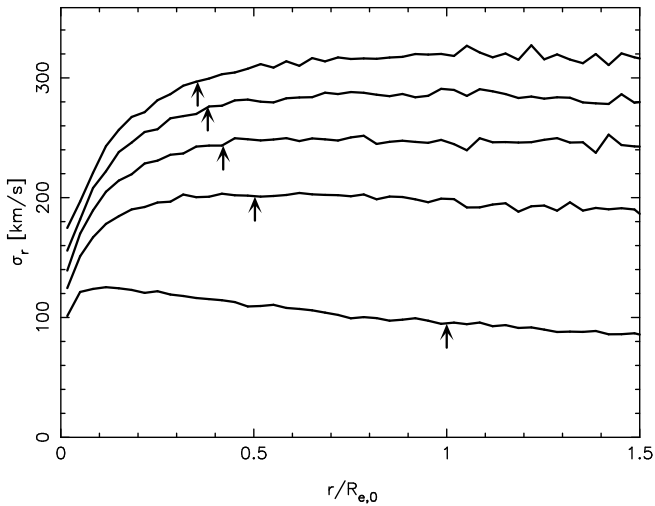}
}
\caption{Evolution of \sig{r}, the radial component of the velocity
  dispersion in spherical coordinates, for the bulge particles in the
  $\rdrei=1$ case.  The profiles are at $\db = 0, 5, 10, 15$ and 20,
  in increasing order.  The arrows indicate \re\ in each
  case.
\label{fig:dispprofs}}
\end{figure}

Growth of the disk compresses the bulge and raises its velocity
dispersion everywhere.  In Figure \ref{fig:dispprofs} we plot examples
of this evolution.  From an initial value of $\sim 100 \kms$, the
bulge in the $\rdrei = 1$ case attains values of $\sig{r} \sim 300
\kms$.  At each output we measure the bulge effective radius, \re, by
computing the circular projected radius containing $50\%$ of the bulge
mass at inclinations $i = 0\degrees, 60\degrees$ and $90\degrees$.  A
detailed analysis of the evolution of the structural parameters will
be presented elsewhere.  We then measured \sig{e} in slits as the
root-mean-square (rms) of the line-of-sight velocity of all particles
within $R_e$
\begin{equation}
\sig{e}^2 = \frac{\sum_{i \in R_e} m_i v_{i,los}^2}{\sum_{i \in R_e} m_i} = \frac{\int_{0}^{\re} I(R) 
(\bar{v}_{los}^2 + \sigma_{los}^2) dR}{\int_0^{\re} I(R) dR},
\label{eqn:apdisp}
\end{equation}
where $\bar{v}_{los}$ is the mean line-of-sight velocity and
$\sigma_{los}$ is the line-of-sight velocity dispersion {\it along the
  slit}, which is placed along the major (\ie\ inclination) axis.  We
repeat the measurements of \sig{e}\ for the same set of inclinations
($i = 0\degrees$, $60\degrees$ and $90\degrees$) and use the average
of these measurements for \sig{e}.  Because \sig{e}\ is not identical
for all viewing orientations, we use the largest difference between
the average \sig{e}\ and the individual values as an estimate of the
uncertainty on \sig{e}.  We use slits of width $20~ \pc$, which is
$\simeq 4\%$ of \rei\ and $\simeq 16\%$ of the smallest $\re$ attained
by the simulations, $\sim 0.14 \kpc$.  The change in \sig{e}\ is not a
result of different radial sampling caused by the change in \re, but
is genuinely caused by evolution of the velocity dispersion, as can be
seen in Figure \ref{fig:dispprofs}.  We also measure \sig{8}, the
velocity dispersion in apertures of \re/8, similarly by restricting
the sum in Eqn. \ref{eqn:apdisp} to that smaller radius.

\begin{figure*}
\centerline{
\includegraphics[angle=-90.,width=0.5\hsize]{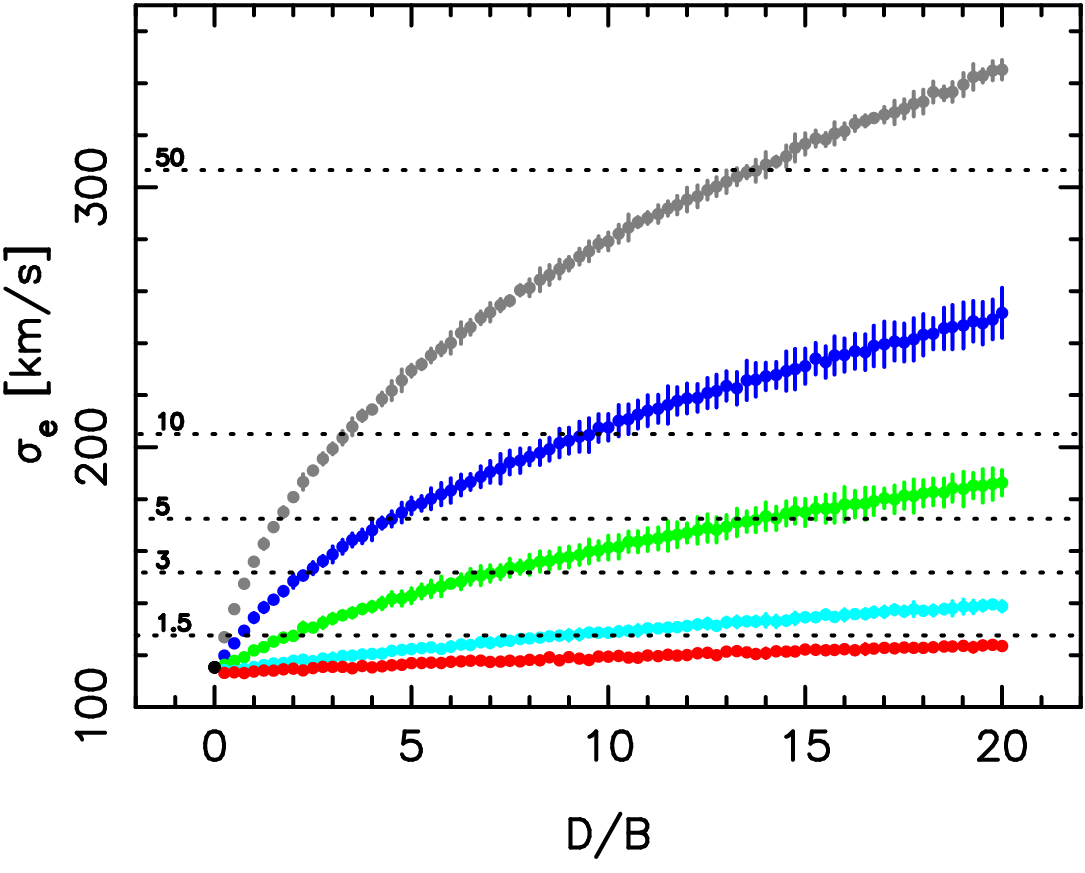}
\includegraphics[angle=-90.,width=0.5\hsize]{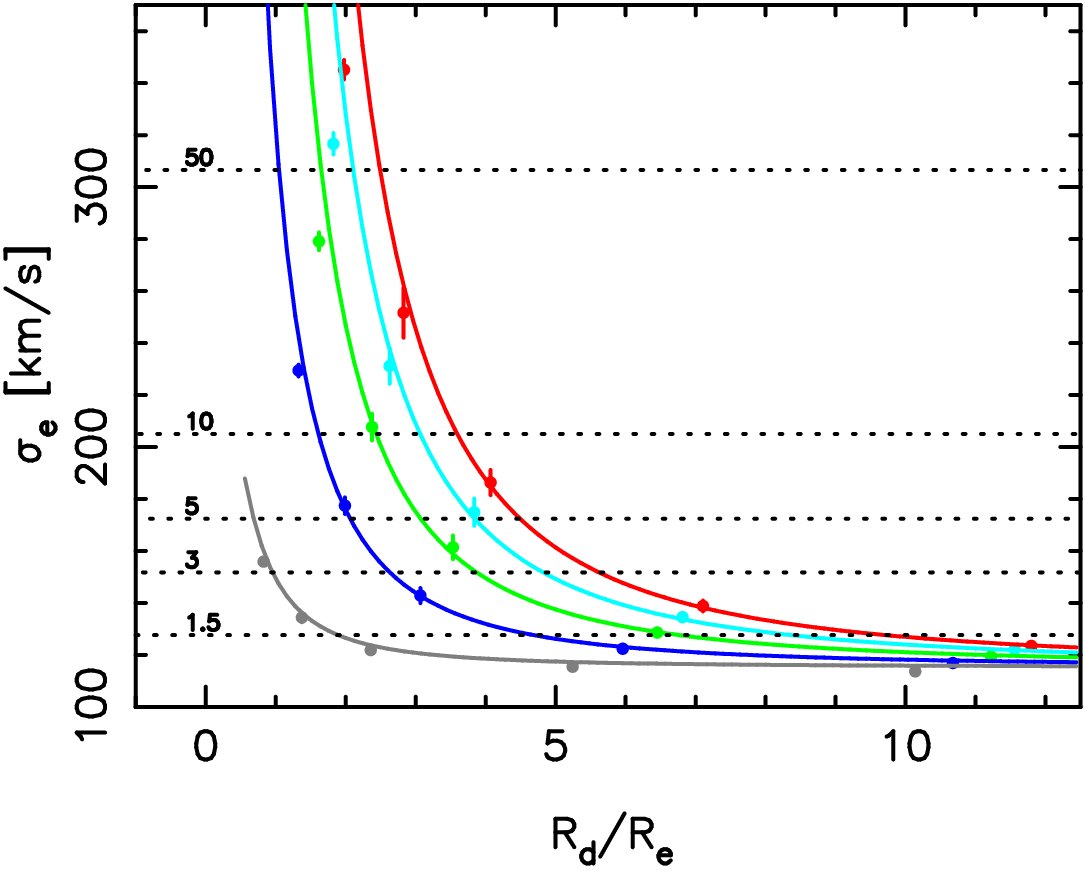}
}
\centerline{
\includegraphics[angle=-90.,width=0.5\hsize]{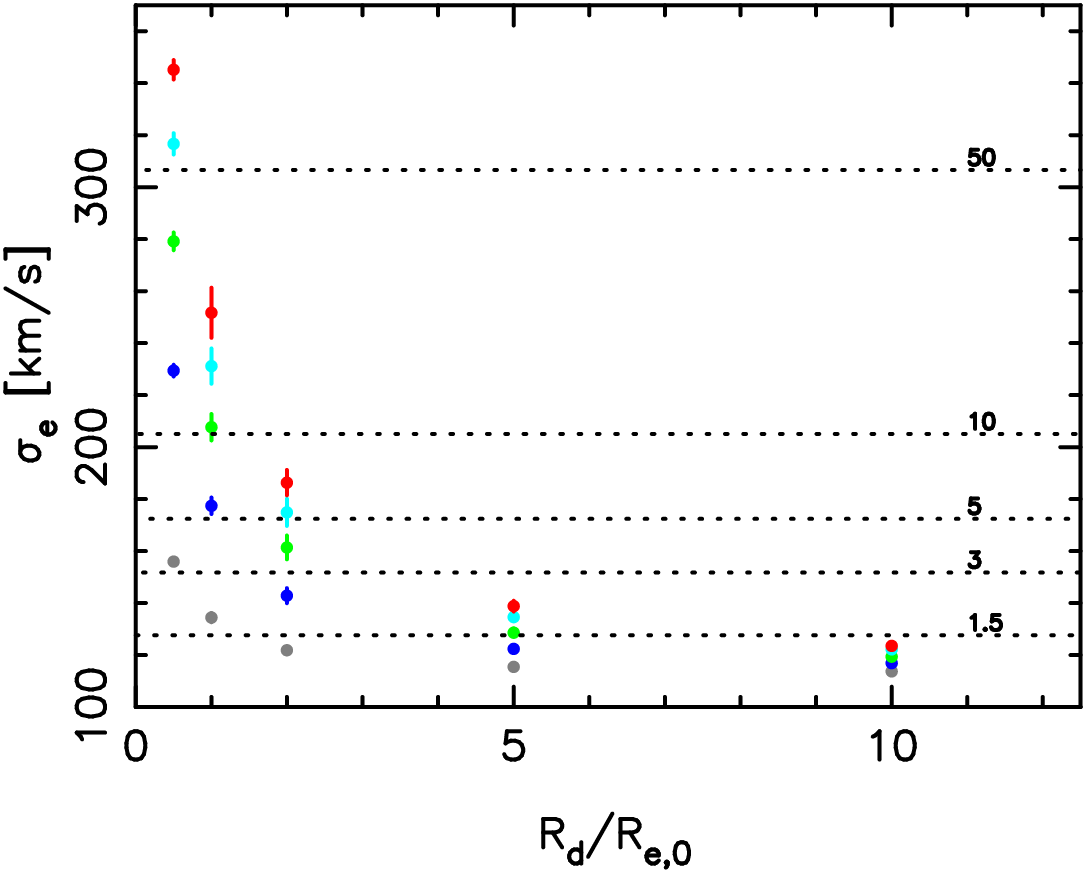}
\includegraphics[angle=-90.,width=0.5\hsize]{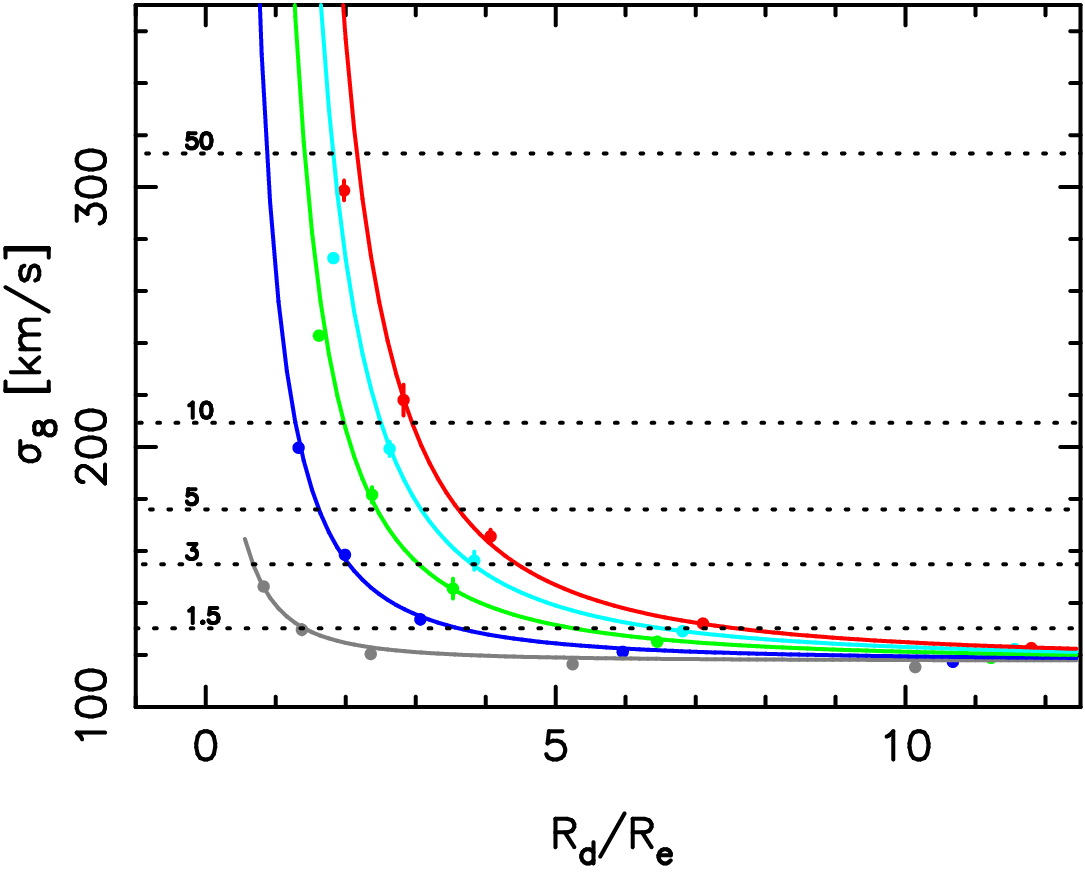}
}
\caption{
{\bf Left}: \sig{e} as a function of \db\ (top) and \rdrei\ (bottom).
Gray, blue, green, cyan and red represent, respectively, \rdrei\ $ =
0.5$, 1, 2, 5 and 10 in the top panel and \db\ = 1, 5, 10, 15 and 20
in the bottom panel.  The initial conditions are indicated by the
black point in the top panel.  {\bf Right}: \sig{e} (top) and \sig{8}
(bottom) as functions of \rdre.  Colors are as in the bottom-left
panel.  The solid lines show our fitting function, Eqn. \ref{eqn:fit}.
In each panel, the dotted horizontal lines show contours of constant
$(\sig{}/\sig{0})^4$, with values indicated above each contour.
\label{fig:sigmas}}
\end{figure*}

The main simulation results are presented in Table
\ref{tab:model_parameters} and plotted in Figure \ref{fig:sigmas}.  In
order to provide a fitting formula to these values, we note that
\citet{wolf+10} find $M_{1/2} \simeq 4 G^{-1} \re
\left<\sig{los}^2\right>$ for pressure-supported systems, where
$M_{1/2}$ is the mass within the half-mass radius and
$\left<\sig{los}^2\right>$ is the luminosity-weighted square of the
line-of-sight velocity dispersion over the entire system.  We
therefore expect \sig{}\ to scale as
\begin{equation} 
\left(\frac{\sig{los,f}(R_f)}{\sig{los,0}(R_0)}\right)^2 = \left(\frac{M_{b,f}(R_f) + M_d(R_f)}{M_{b,0}(R_0)}\right) \left(\frac{R_0}{R_f}\right)
\label{eqn:dsig} 
\end{equation} 
where subscripts $f$ and $0$ indicate final and initial values and
$M_b(R)$ and $M_d(R)$ indicate bulge and disk masses within radius
$R$.  Integrating over radius, we expect
\begin{eqnarray} 
\frac{\sig{e}}{\sig{e,0}} & = &
\left(1+ 2\gamma\left[ 1 - \left(1+\frac{\re}{\rd}\right)e^{-\rerd} \right] \frac{D}{B} \right)^\delta \nonumber \\
& \equiv & {\cal F}
\label{eqn:fit} 
\end{eqnarray} 
where $\sig{e,0}$ is the initial dispersion of the bulge.  We stress
that ${\cal F}$ is defined to be a {\it photometric}, not kinematic,
quantity.  Note that at fixed disk-to-bulge mass ratio, $\db$, the
minimum in ${\cal F}$ occurs at $\rerd = 0$, \ie\ as $\rd \rightarrow
\infty $.  In deriving Eqn.  \ref{eqn:fit} from the more general Eqn.
\ref{eqn:dsig} we have assumed that the disk is exponential; this is
true for our simulations, but need not be the case in nature
\citep[e.g.][]{boe_etal_03, dutton_09}.  If $\gamma = 1$ then the term
in the outer brackets on the right hand side of Eqn.  \ref{eqn:fit} is
merely the ratio of final (bulge$+$disk) mass to initial (bulge only)
mass within \re.  (We ignore the dark matter halo in this calculation
since the dark-to-bulge mass fraction within \rei\ is less than 2\%.)
We have also assumed that the disk scale-height is small compared to
the effective radius of the bulge, but the factor $\gamma$ is
introduced to account for some of deviations resulting from this
assumption.  We have neglected the compression of the bulge in
deriving this expression, \ie\ we assume that $\re = \rei$, which
Table \ref{tab:model_parameters} clearly shows is not the case.  We
fold the uncertainty resulting from this assumption into the free
parameter $\delta$, which would be $0.5$ if $\re=\rei$.  The best fit
for $2 \leq \rdre \leq 9$, $\db > 0$ and $1 \leq \sig{e}/\sig{e,0}
\leq 2$ is $\gamma = 0.3$ and $\delta = 1.76$ which gives a $\chi^2 =
172$ for 196 data points.  The best fit value with $\gamma = 1$ is
$\delta = 0.64$ with $\chi^2 = 562$.  Likewise, we fitted best-fit
parameters for \sig{8} obtaining $\gamma = 0.02$ and $\delta = 15.92$
with a $\chi^2 = 719$.  These best fits are shown in Figure
\ref{fig:sigmas}.  Because the compression of the bulge depends only
on the ratios of bulge-to-disk masses and sizes, Eqn.  \ref{eqn:fit}
remains true for any galaxy or bulge mass, \ie\ for any \sig{e,0}.

\begin{figure}
\includegraphics[angle=-90.,width=\hsize]{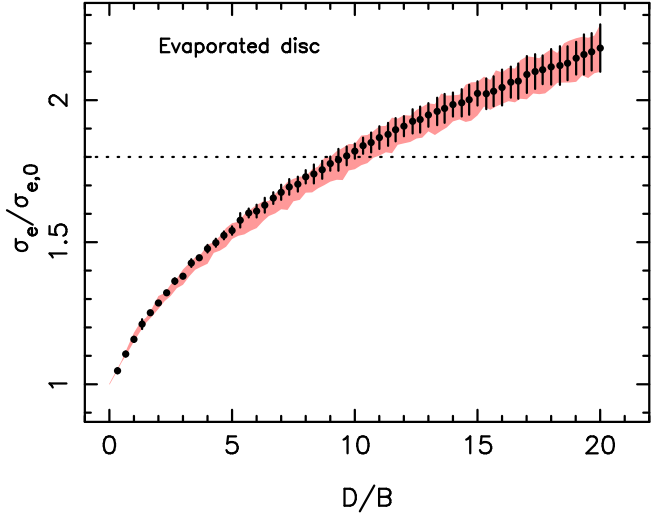}
\caption{Evolution of \sig{e}\ in the $\rdrei=1$ model as the disk is
  grown (red shaded region) and subsequently evaporated (black points
  with error bars).  The dotted horizontal line indicates
  $\sig{e}/\sig{e,0} = 1.8$ for which $\dMbh = 10.5$ if $\beta =
    4$.
    \label{fig:evap}}
\end{figure}

In order to check whether the bulge and halo respond adiabatically to
the growth of the disk we slowly evaporated the disk from the final
state of the simulation with $\rdrei = 1$. The results of this test
are shown in Figure \ref{fig:evap} and indicate that indeed the
response is adiabatic to good approximation.

\subsection{Other dependencies and sources of scatter}

\begin{figure}
\includegraphics[angle=-90.,width=0.9\hsize]{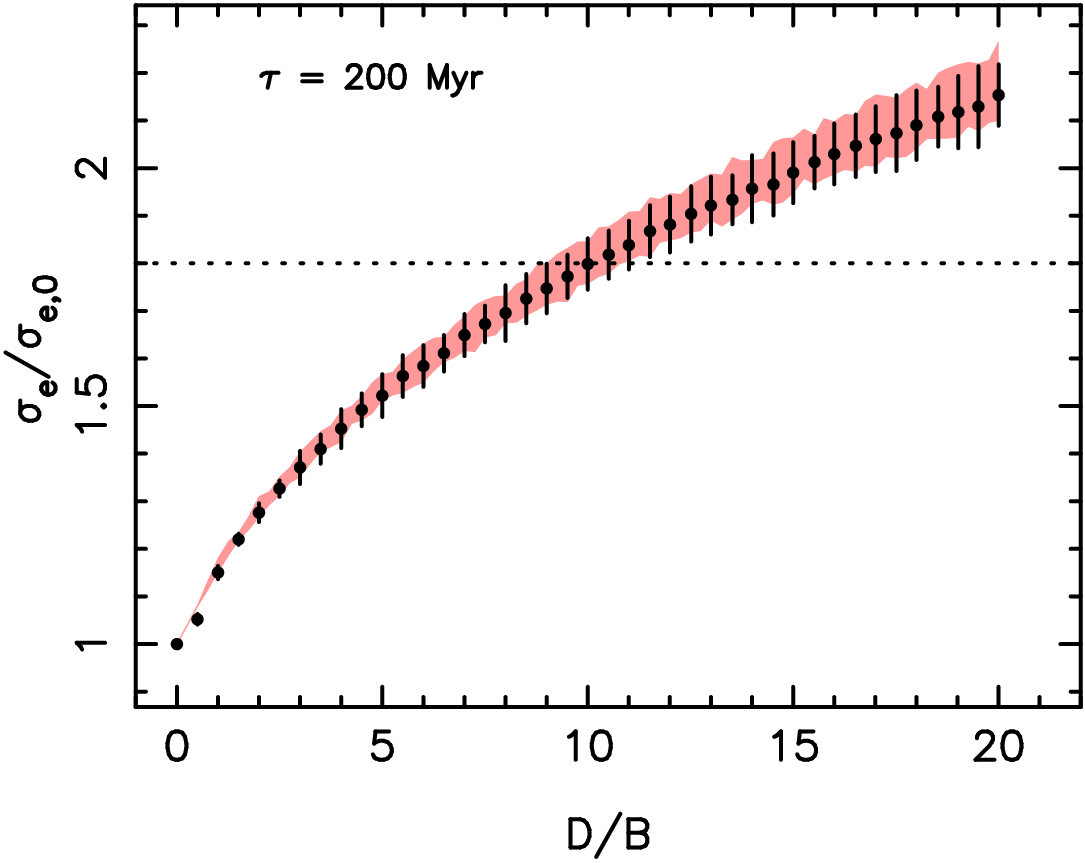}
\includegraphics[angle=-90.,width=0.9\hsize]{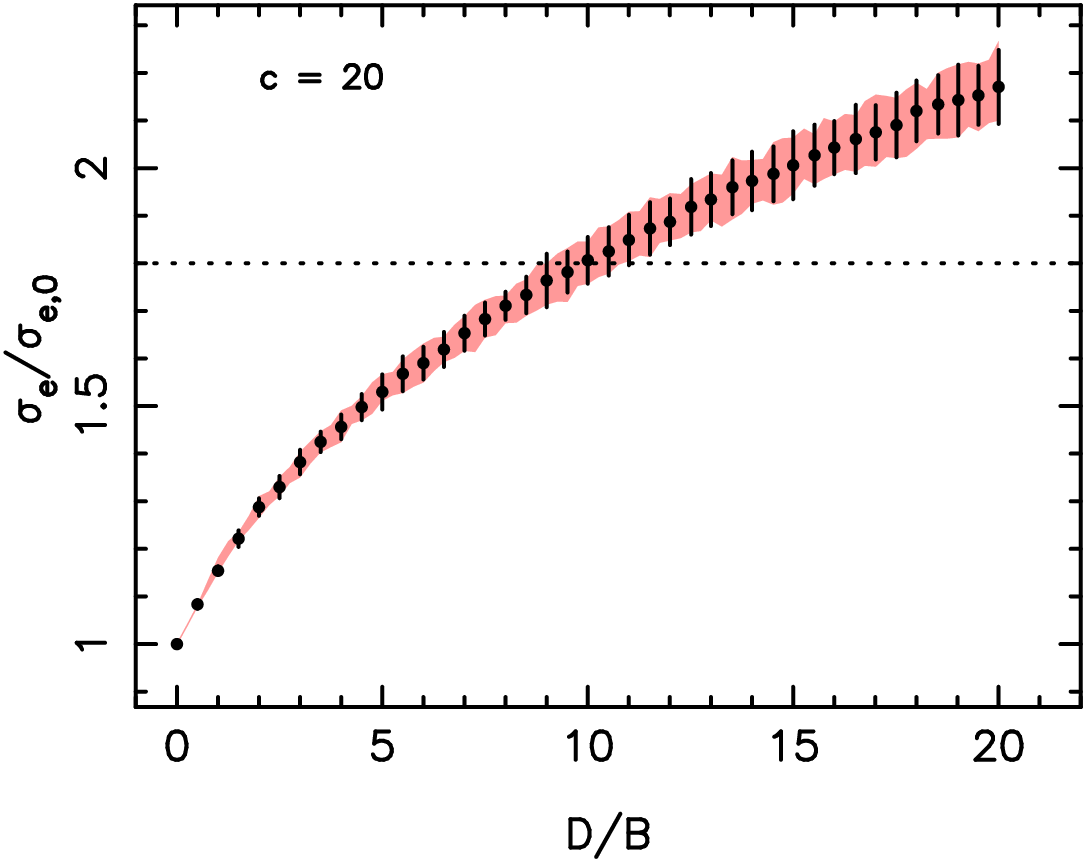}
\caption{Evolution of \sig{e}\ in the $\rdrei=1$ model for different
  variations from our fiducial simulation.  In both panels the shaded
  red region represents the evolution in the fiducial case, while the
  black points with error bars show the variant simulation results.
  The dotted horizontal lines indicate $\sig{e}/\sig{e,0} = 1.8$ for
  which $\dMbh = 10.5$.  Top: Effect of growing the disk in 200 Myr
  instead of 2 Gyr.  Bottom: Effect of the halo having concentration
  $c=20$ instead of $c=10$.
\label{fig:variants}}
\end{figure}

Eqn. \ref{eqn:fit} allows us to estimate the increase in \sig{e}\ for
a bulge given a galaxy's photometric decomposition.  We now explore
the amount of scatter that can occur in these estimates for a given
observed density distribution.  Some of the effects we consider here
will compress the bulge to a different extent, so both \sig{e}\ and
\re\ will change; it may happen however that Eqn. \ref{eqn:fit} still
provides a good approximation to the evolution of \sig{e}.  Because we
use Eqn. \ref{eqn:fit} to estimate the offset of galaxies from the
\Msig\ relation, we are primarily interested in those changes which
Eqn. \ref{eqn:fit} does {\it not} reproduce, and we consider this to
be the scatter of interest here.

The (re)-assembly of disk galaxies is not necessarily a slow,
adiabatic process.  A possible source of scatter might therefore be
due to disks growing more rapidly than assumed here.  In order to test
what the effect of faster disk growth may be, we grow the disk ten
times faster, \ie\ within 200 Myr.  The results are shown in the top
panel of Figure \ref{fig:variants}.  The effect of the different
growth rate is negligible.

At a given mass, the concentration of dark matter halos can vary
substantially \citep[e.g.][]{wechsler+02}.  We explore what effect
this might have on \sig{e}\ by re-running our simulation with a halo
having $c=20$.  The bottom panel of Figure \ref{fig:variants} shows
that \sig{e}\ is barely changed, undoubtedly because the galaxy is
baryon-dominated in the bulge region.  Reasonable variations in halo
concentration therefore do not produce any significant scatter in the
\Msig\ relation.

\begin{figure}
\includegraphics[angle=-90.,width=\hsize]{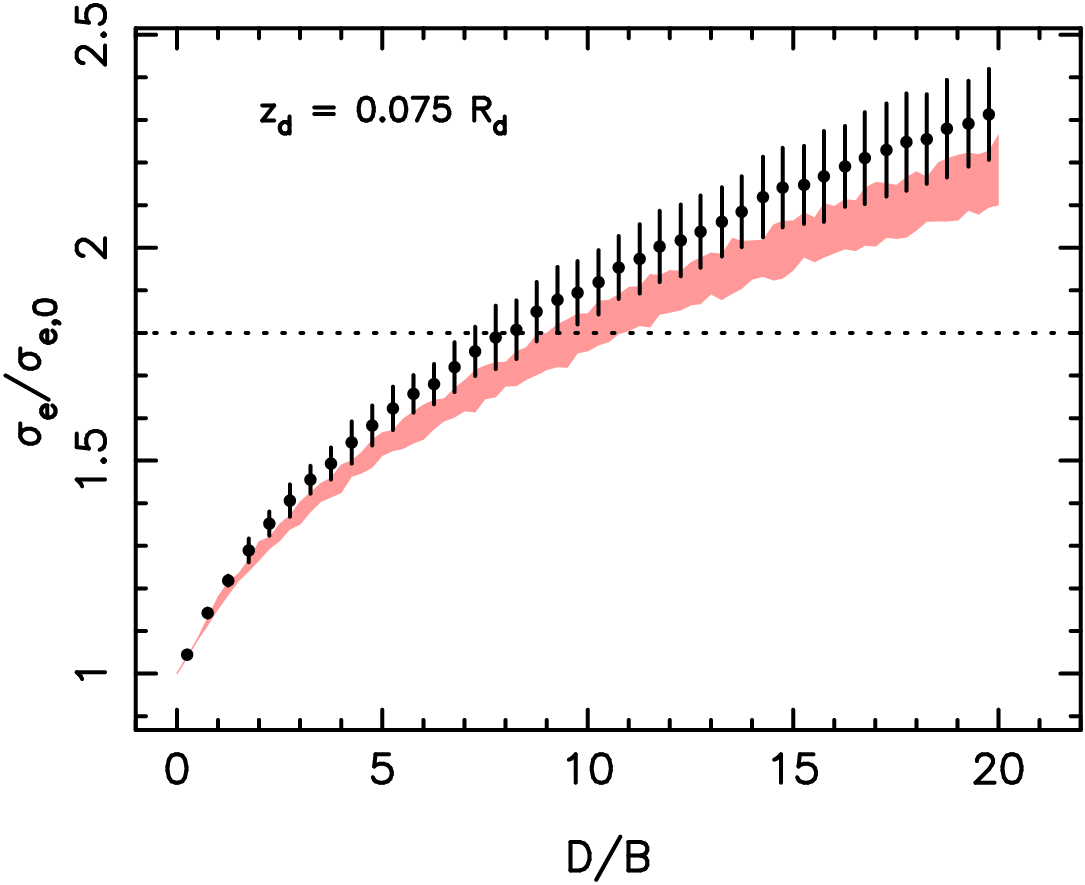}
\includegraphics[angle=-90.,width=\hsize]{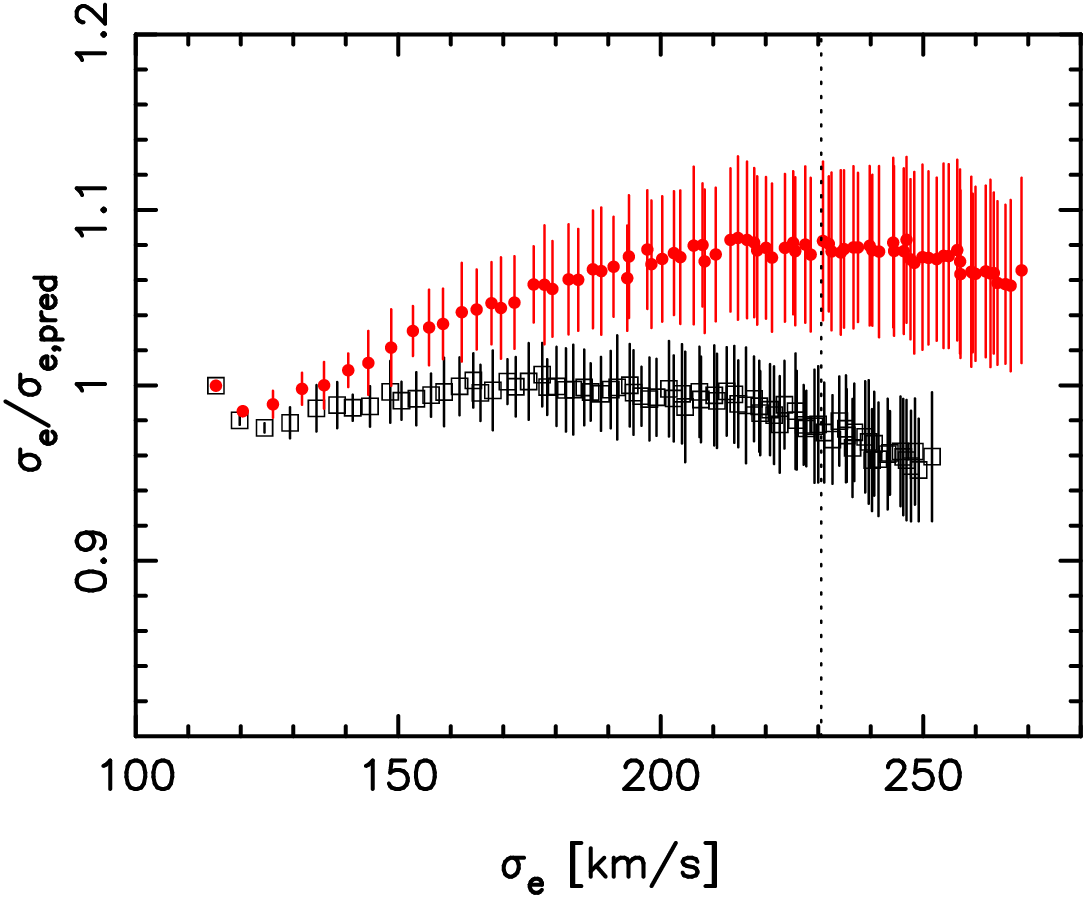}
\caption{
Comparision of the evolution of \sig{e}\ in the $\rdrei = 1$ model
with a thinner disc $\zd = 0.075\rd$ instead of the fiducial $\zd =
0.15\rd$.  Top: The shaded red region represents the evolution in the
fiducial case, while the black points with error bars show the thinner
disk.  The dotted horizontal line indicates $\sig{e}/\sig{e,0} = 1.8$
for which $\dMbh = 10.5$.  Bottom: A comparison of the fit of Eqn.
\ref{eqn:fit} for the fiducial case (black open squares) and with $\zd
= 0.075\rd$ (red filled circles).  The dotted vertical line indicates 
$\sig{e}/\sig{e,0} = 1.8$.
\label{fig:thickness}}
\end{figure}

All simulations above used $\zd = 0.15\rd$.  In the top panel of
Figure \ref{fig:thickness} we show the effect of halving \zd.  This
increases the final \sig{e} to $268.7 \pm 13.3 \kms$, an increase by
6.8\% while \re\ decreases by $\sim 3.3\%$.  The bottom panel of
Figure \ref{fig:thickness} shows that, taken together, these
differences lower the quality of the fit of Eqn.  \ref{eqn:fit} for
$\zd = 0.075\rd$ compared with that for the standard $\zd = 0.15\rd$,
although all cases still have errors of less than $15\%$.  At
$\sig{e}/\sig{e,0} = 1.8$ the {\it maximum} error is about 12\%, which
we adopt as our estimate for the scatter due to disk thickness.

\begin{figure}
\centerline{
\includegraphics[angle=-90.,width=\hsize]{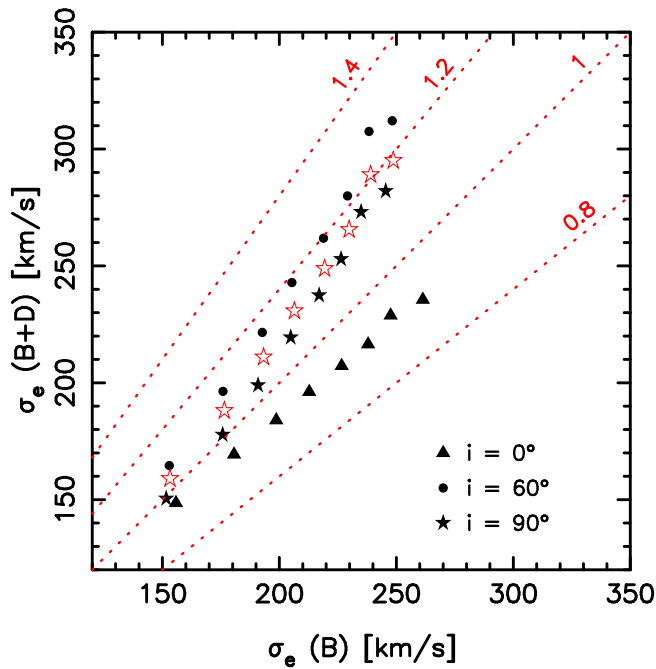}
}
\caption{The effect of including disk stars in the measurement of
  \sig{e}.  \sig{e}(B) represents the measurement from bulge particles
  only while \sig{e}(B+D) includes disk particles in the measurement.
  Dotted lines have constant slope, as indicated along each line.  The
  different points represent the effect of disk contamination as \db\
  increases from 2.5 to 20.  Different filled symbols correspond 
  to different galaxy inclinations, as indicated.  Open (red) stars 
  correspond to inclination-averaged values.
  \label{fig:diskcont}}
\end{figure}

Another source of scatter comes from the contamination of measured
bulge kinematics by the kinematics of the disk, which is almost
inevitable in real galaxies.  Exploring this effect requires that we
set up equilibrium kinematics for the disks in the $\rdrei = 1$ model
at various values of \db.  We set the kinematics of the disks to give
constant Toomre-$Q = 1.5$, as described in \citet{deb_sel_00}.  For
this we calculate the potential using a hybrid polar-grid code with
the disk on a cylindrical grid and the bulge$+$halo on a spherical
grid \citep{sellwo_03}.  Figure \ref{fig:diskcont} shows the effect of
disk contamination: changes in \sig{e} can be either positive or
negative, but generally $|\sig{e}(B)-\sig{e}(B+D)|/\sig{e}(B)\la
25\%$.  The error increases with \sig{e}, which is a result of the
increasing \db. An independent analysis of the effect of disk
contamination on \sig{e} by Hartmann et al. ({\it in preparation})
also finds fractional changes $\la 25\%$.  The open stars in Figure
\ref{fig:diskcont} show the inclination-averaged values of
$\sig{e}(B)$ versus $\sig{e}(B+D)$; the differences between the means
are generally less than 20\%.  However on average $\sig{e}(B+D)$ is
systematically larger than $\sig{e}(B)$.

Based on these tests we conclude that there is $\la 30\%$ uncertainty
in the degree to which classical bulges are compressed.

\subsection{Observational evidence for bulge compression}

We now present evidence that bulge compression associated with disk
regrowth has occurred in nature by comparing the properties of
classical bulges and elliptical galaxies.  This requires a large
sample of disk galaxies with bulge$+$disk decompositions.
\citet{gadotti09} presented a detailed structural analysis of nearly
1000 galaxies from the Sloan Digital Sky Survey (SDSS) \citep{sdss},
classifying them into ellipticals or disks, distinguishing the latter
by whether they host classical or pseudo bulges.
\citet{gadotti_kauffmann_09} present the velocity dispersions within
$\re/8$, hereafter \sig{8}, for a fraction of these galaxies.  We use
these data to compare the distributions of \sig{8}\ for ellipticals
and classical bulges.  The sample contains 196 elliptical galaxies and
176 unbarred classical bulges with kinematic data.  \citet{gadotti09}
classifies the bulges based on the Kormendy (mean effective surface
brightness $\left<\mu_e\right>$ versus \re) relation
\citep{kormendy77}.  \citet{fis_dro_08} identify S\'ersic index $n=2$
as the dividing line between pseudo and classical bulges, with the
latter having $n>2$.  We apply this additional criterion to the
sample, which leaves 166 galaxies as our final sample of unbarred,
classical bulges.

\begin{figure}
\centerline{
\includegraphics[angle=-90.,width=\hsize]{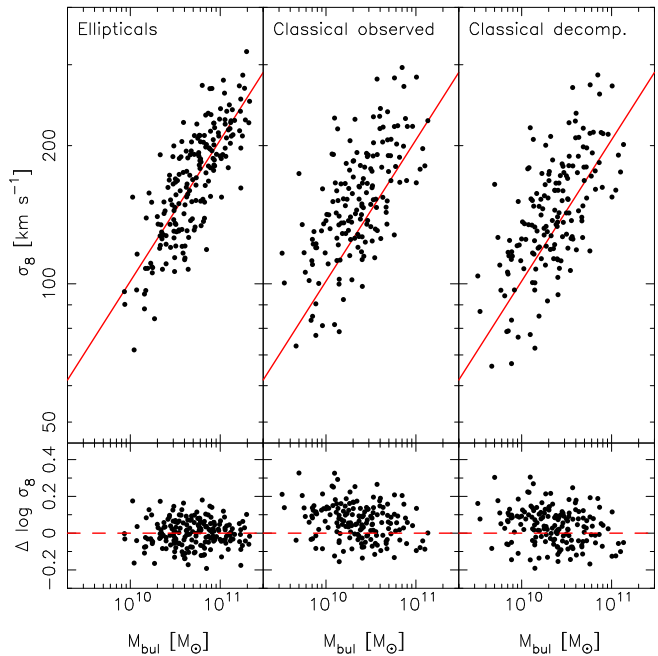}}
\caption{Distributions of ellipticals and classical bulges, taken from
  \citet{gadotti09} and \citet{gadotti_kauffmann_09}, in the
  \sig{8}-\Mbul\ plane. Left: Elliptical galaxies. Center: Classical
  bulges in unbarred galaxies.  Right: The same sample as in the
  central panel, but using the decompressed values for \sig{8}
  obtained by applying Eqn. \ref{eqn:fit}.  The (red) solid lines in
  each of the upper panels show the best fit to the elliptical
  galaxies.  The bottom panels show the residuals for each sample from
  this best fit to the ellipticals: $\Delta \log \sig{8} = \log
  \sig{8} - \log \sig{8,{\rm fit}}$, where $\sig{8,{\rm fit}}$ is
    the \sig{8}\ value from the fit to the ellipticals of a given
    mass.
    \label{fig:gadotti}}
\end{figure}

Figure \ref{fig:gadotti} plots the distribution of galaxies in the
\sig{8}-\Mbul\ plane for ellipticals, observed unbarred classical
bulges, and the same classical bulges if they are decompressed using
Eqn. \ref{eqn:fit}.  We obtain $\Mbul$, \db\ and \rerd\ from the the
exponential disk$+$ S\'ersic bulge decompositions of \citet{gadotti09}
and \sig{8}\ from \citet{gadotti_kauffmann_09}.  We decompress to
obtain \sig{8,0}\ using Eqn. \ref{eqn:fit} fitted to \sig{8}.

\begin{figure}
\centerline{
\includegraphics[angle=-90.,width=\hsize]{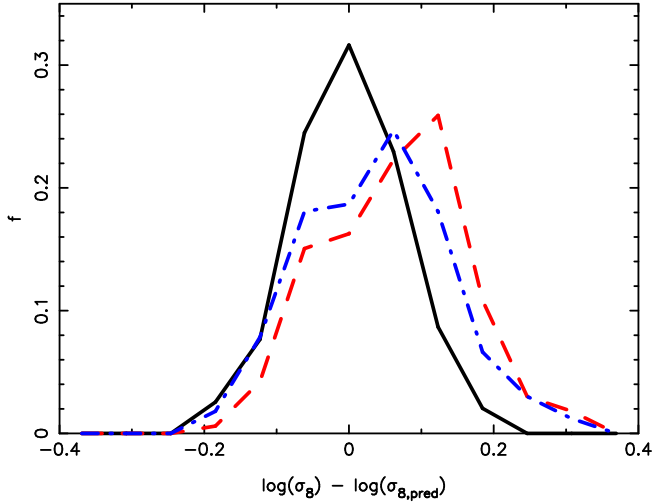}
}
\caption{Distributions of the residuals shown in the lower panels of
  Fig. \ref{fig:gadotti}. The solid (black) line is for the
  ellipticals, the dashed (red) line is for observed classical bulges
  (as described in the text) and the dot-dashed (blue) line is for the
  classical bulges decompressed using Eqn. \ref{eqn:fit}.
\label{fig:gadottiresids}}
\end{figure}

\begin{figure}
\centerline{
\includegraphics[angle=-90.,width=\hsize]{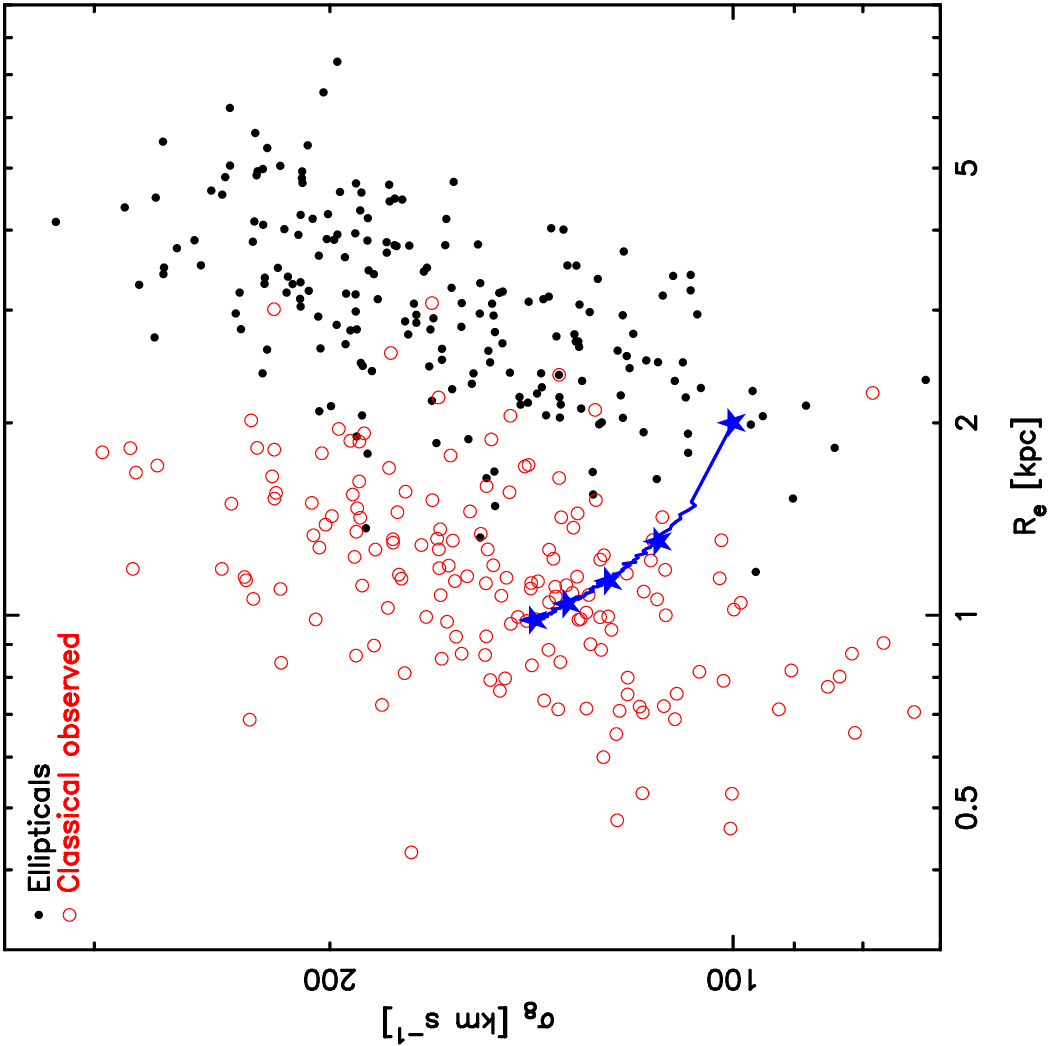}
}
\caption{The \sig{8}-\re\ projection of the fundamental plane for the
  \citet{gadotti_kauffmann_09} sample.  The (black) filled circles
  show elliptical galaxies while the (red) open circles show the
  observed unbarred classical bulges.  The diagonal (blue) line shows
  the evolution of the model in the simulation with $\rdrei = 2$,
  assuming $\rei = 2$ kpc and $\sig{e,0} = 100 \kms$.  The star
  symbols correspond to the system at $\db = 0$ (bottom right), 5, 10,
  15 and 20 (top left).
\label{fig:fundplane}}
\end{figure}

The line in the top panels of Figure \ref{fig:gadotti} shows the fit
to the ellipticals: $\sig{8}/\kms = (\Mbul/3051\Msun)^{0.30783}$.  The
observed classical bulges have larger \sig{8}, on average, than the
ellipticals at a given \Mbul.  When we decompress the bulges their
offset from the elliptical relation is significantly reduced, as can
be seen in the bottom panels of Figure \ref{fig:gadotti}. The
distributions of the residuals from the fit to ellipticals are shown
in Figure \ref{fig:gadottiresids}. The means of the residuals are 0.06
dex and 0.04 dex for the observed and decompressed classical bulges,
respectively.  A two-sample unbinned K-S test comparing the
ellipticals and bulges shows that the probability that the residuals
are drawn from the same distribution is $3\times10^{-9}$ for the
observed bulges and a much larger, though still formally small,
$3\times10^{-6}$ for the decompressed bulges.  In Figure
\ref{fig:fundplane} we plot the distribution of the bulges and
ellipticals in the \sig{8}-\re\ plane.  As was also found by
\citet{gadotti09}, the observed classical bulges are offset to larger
\sig{8}\ and smaller \re\ relative to the ellipticals in this
projection of the Fundamental Plane, as expected if bulges are
compressed by disks.

While Figures \ref{fig:gadotti} and \ref{fig:gadottiresids} do provide
evidence for bulge compression, they also show that the \sig{8}\ is
even larger than predicted by our simple model.  Possibly this is
because disks are more concentrated than exponential at the center, as
proposed by \citet{boe_etal_03} and \citet{dutton_09}.  Alternatively,
contamination of \sig{8}\ by the disk, or differences in formation
histories, could be to blame for (part of) the offset between
ellipticals and classical bulges.  We note that if disks are more
concentrated than exponential then their effect is to further compress
bulges and we have under-estimated the evolution of \sig{e}.
Exploration of this issue is deferred to a future publication.

%%%%%%%%%%%%%%%%%%%%%%%%%%%%%%%%%%%%%%%%%%%%%%%%%%%%%%%%%%%%%%%%%%%%%

\section{Consequences for the \Msig\ Relation}
\label{sec:msig}

The steepness of the \Msig\ relation implies that the maximum factor
of $\sim 3$ increase in \sig{e}\ obtained in the simulations would
require a factor of $\sim 80$ increase in \Mbh\ for the SMBH to remain
on the relation, more than $6\times$ larger than the factor of 20 by
which the stellar mass grew.  We define the factor by which \Mbh\ must
grow to remain on the \Msig\ relation, $\dMbh \equiv
M_{\bullet,f}/M_{\bullet,0} = (\sig{e}/\sig{e,0})^\beta$, where
subscripts $0$ and $f$ indicate initial and final values.  The dotted
contours in Figure \ref{fig:sigmas} indicate \dMbh\ assuming $\beta =
4$.  Values of \dMbh\ for the simulations are listed in Table
\ref{tab:model_parameters}.  In general $\dMbh > 3$ requires that
$\rdre \la 5$ and $\db \ga 2$.

\subsection{Evolution of slope and zero-point}

\begin{figure}
\centerline{
\includegraphics[angle=-90.,width=\hsize]{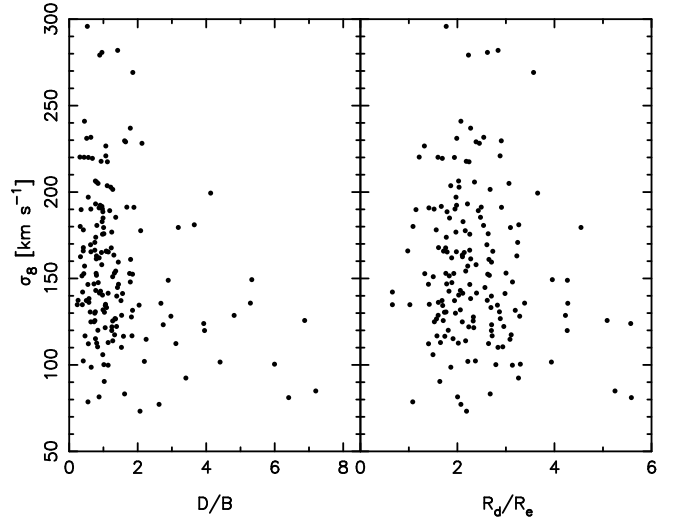}
}
\caption{Dependence of \sig{8}\ on the parameters of bulge$+$disc
  decompositions for unbarred classical bulges in the sample of
  \citet{gadotti09}.\label{fig:gadottistruct}}
\end{figure}

In Figure \ref{fig:gadottistruct} we plot the distribution of \sig{8}\
as a function of \db\ and of \rdre\ for the \citet{gadotti09} sample;
a weak correlation between \db\ and \sig{8} is present (Spearman
$r_{sp} = -0.22$, Kendall $\tau = -0.15$), which is statistically
significant at less than $3\sigma$.  The correlation between \rdre\
and \sig{8}\ is even weaker (Spearman $r_{sp} = -0.10$, Kendall $\tau
= -0.07$).  We therefore neglect these weak correlations.  If $
\sig{e}/\sig{e,0} = {\cal F}$, and the average \sig{e}/\sig{e,0} at a
given \sig{e,0} is $\left<{\cal{F}}\right>$, then neglecting these
correlations implies that $\left<{\cal{F}}\right>$ is independent of
\sig{e,0}.  Then we can write the \Msig\ relation for compressed
bulges, if \Mbh\ does not change while disks grow, as
\begin{equation}
\log \Mbh = \alpha - \beta \log\left<{\cal{F}}\right> + \beta \log \sig{e}
\end{equation}
\ie\ the slope of the relation remains $\beta$ but the zero-point changes
by
\begin{equation}
\delta\alpha = - \beta \log \left<{\cal{F}}\right>.
\label{eqn:zpoint}
\end{equation}
Because ${\cal{F}} \ge 1$ (compression can only increase \sig{e}) Eqn.
\ref{eqn:zpoint} implies that $\delta\alpha < 0$.  Therefore, if \Mbh\
does not grow during disk formation, the \Msig\ relation of compressed
bulges will be parallel to, but offset below, the \Msig\ relation for
elliptical galaxies.  Failure to find such an offset would strongly
suggest that SMBHs grow along with disks.

\subsection{Predicted offset from photometric samples}

\begin{figure*}
\centerline{
\includegraphics[angle=0.,width=0.5\hsize]{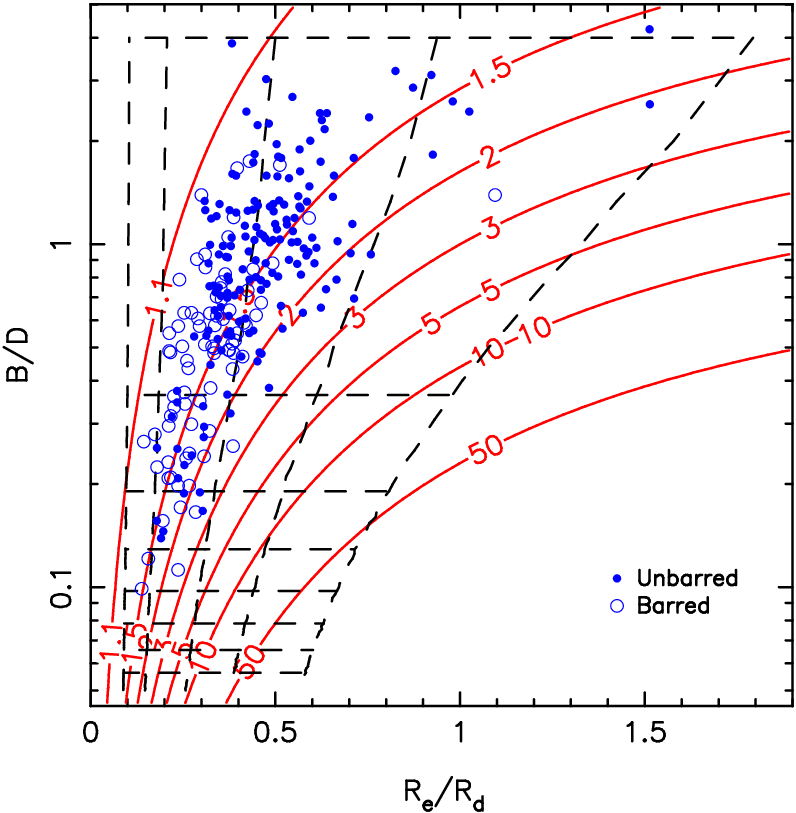}
\includegraphics[angle=0.,width=0.5\hsize]{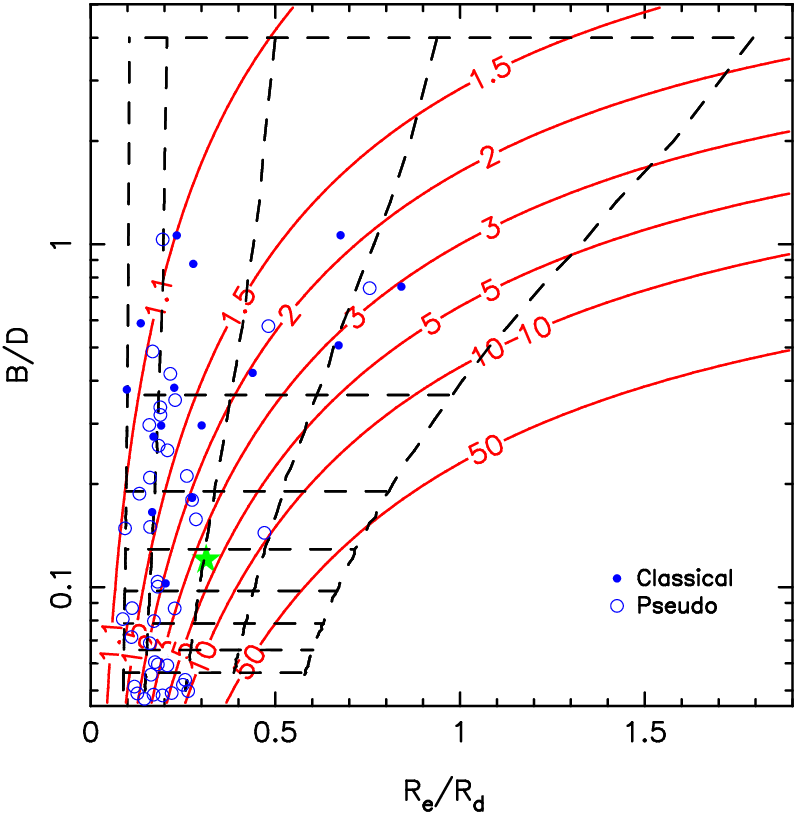}
}
\caption{Contours of predicted \dMbh\ in the \bd-\rerd\ plane assuming
  the \citet{gultekin+09} value of $\beta = 4.24$ (solid lines).  The
  dashed lines correspond to our simulation grid, with horizontal
  lines at fixed \db\ and roughly vertical lines at constant \rdrei.
  A galaxy growing a disk at fixed \rd\ evolves from top to bottom
  parallel to the dashed lines.  Left: The (blue) circles are data
  from \citet{gadotti09}, with open and filled circles corresponding
  to barred and unbarred galaxies, respectively.  See text for details
  of how the parameters for the barred galaxies are computed.  Right:
  The (blue) circles are data from \citet{graham_03}, with open and
  filled circles corresponding to galaxies with $n<2$ (pseudo) bulges
  and $n>2$ (classical) bulges, respectively.  The (green) star
  represents the Milky Way based on the model of \citet{bis_ger_02}.
\label{fig:deltaMs}}
\end{figure*}

In order to provide a quantitative prediction for the change in the
zero-point we use two samples of galaxies with detailed photometric
decompositions.  The first is the sample of \citet{gadotti09}, for
which each galaxy is fit with three components; a bulge, a disk, and,
where necessary, a bar.  We split the sample by whether the galaxy is
barred or not.  In our analysis, for the barred galaxies we treat the
bar as part of the disk when computing \db\ values and use \rd\ from
the disk, not the bar.  Because bars are at the centers of galaxies,
our assumption underestimates the fraction of disk$+$bar mass within
the bulge effective radius, and therefore also
$\left<{\cal{F}}\right>$.  The second sample is the complete and
volume-limited catalogue of 86 low-inclination disk galaxies of all
Hubble types observed by \citet{dej_vdk_94}.  For 75 of these,
\citet{graham_03} fitted S\'ersic bulge$+$exponential disk
decompositions in the $K$-band, regardless of whether they are barred
or not.  We select classical bulges from this sample as those galaxies
having $n>2$, leaving us with 15 galaxies.

\begin{figure}
\centerline{
\includegraphics[angle=-90.,width=\hsize]{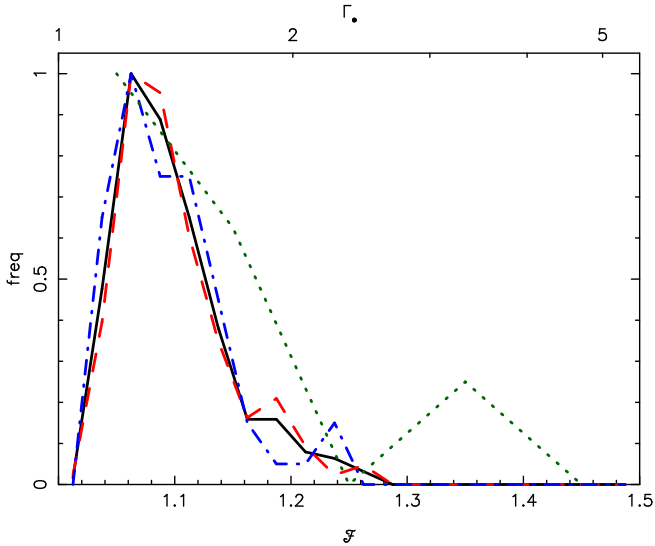}
}
\caption{The distribution of ${\cal{F}}$ for the various photometric
  samples.  The dotted (green) line shows the classical bulges in the
  \citet{graham_03} sample. The remaining lines are for the
  \citet{gadotti09} sample of classical bulges: the dashed (red) line
  is for unbarred galaxies, dot-dashed (blue) line for barred galaxies
  and solid (black) line for all galaxies.  For ease of comparison,
  all distributions have been normalized to unit peak value.  The top
  border is labelled by \dMbh\ assuming $\beta = 4.24$
  \citep{gultekin+09}.
\label{fig:ffactor}}
\end{figure}

Figure \ref{fig:deltaMs} plots the distribution of both samples in the
\rdre-\db\ plane and overlays contours of \dMbh.  All but one of the
classical bulges in both samples have $\db < 10$, whereas many of the
pseudo bulges in the \citet{graham_03} sample have $\db > 10$.  The
majority of the galaxies in the \citet{gadotti09} sample cluster in
the range $0.2 \la \rerd \le 0.6$ ($1.6 \le \rdre \la 5$).  For about
half of all galaxies $\dMbh > 1.4$, while a small fraction ($\sim
33\%$ of the \citet{graham_03} sample and $\sim 8\%$ in the
\citet{gadotti09} sample) has $\dMbh > 2$.  A small number of galaxies
fall outside the simulation grid.  In calculating \dMbh\ for these
galaxies, we extrapolate Eqn. \ref{eqn:fit} to outside our simulation
grid.  These are mostly however at large \db, and populated solely by
pseudo bulges, rather than classical ones.  Many more galaxies in the
\citet{gadotti09} sample have $\db<1$ than in the \citet{graham_03}
sample.  The difference cannot be attributed to the different
photometric decompositions since the same difference is present also
for unbarred galaxies in the \citet{gadotti09} sample.  Therefore
together these two samples should give some indication of the
uncertainty in the photometric parameter $\left<{\cal{F}}\right>$.
The distributions of ${\cal{F}}$ for both samples are shown in Figure
\ref{fig:ffactor} and the results listed in Table \ref{tab:ffactors}.
Notwithstanding the differences between the samples, we find a narrow
range of $1.096 \leq \left<{\cal{F}}\right> \leq 1.122$.  Assuming
$\beta=4.24$ \citep{gultekin+09}, this corresponds to $1.48 \leq
\left<\dMbh\right> \leq 1.63$ and $-0.21 \leq \delta\alpha \leq
-0.17$.  For their full sample \citet{gultekin+09} measured $\alpha =
8.12 \pm 0.08$; with such a small uncertainty on $\alpha$, offsets
between classical bulges and elliptical galaxies in the \Msig\
relation should be measureable.  We estimated above a scatter due to
modelling uncertainties of order $30\%$, but the effect we are looking
for here is systematic, so it should be detectable if present.

\begin{table}[!ht]
\begin{centering}
\begin{tabular}{rcc} \\ \hline
\multicolumn{1}{c}{Sample} &
\multicolumn{1}{c}{$N_g$} &
\multicolumn{1}{c}{$\left<{\cal{F}}\right>$} \\ \hline
  Unbarred \citet{gadotti09} & 166 & $1.098 \pm 0.004 $ \\
    Barred \citet{gadotti09} &  80 & $1.096 \pm 0.005 $ \\
       All \citet{gadotti09} & 246 & $1.098 \pm 0.003 $ \\
 Classical \citet{graham_03} &  15 & $1.122 \pm 0.025 $ \\ \hline
\end{tabular}
\caption{Values of $\left<{\cal{F}}\right>$ for different photometric
  samples.  $N_g$ indicates the number of galaxies in each sample.
  The error on $\left<{\cal{F}}\right>$ is purely statistical. 
\label{tab:ffactors}}
\end{centering} 
\end{table}

%%%%%%%%%%%%%%%%%%%%%%%%%%%%%%%%%%%%%%%%%%%%%%%%%%%%%%%%%%%%%%%%%%%%%

\section{Testing for Offsets}
\label{sec:offsets}

We now test for offsets between the \Msig\ relations of elliptical
galaxies and of classical bulges.  We show that there is no
significant offset between the two populations.  If we decompress the
bulges using Eqn. \ref{eqn:fit} then a small but significant offset
occurs, which supports our claim that an offset should have been
detected if SMBHs had not grown in mass along with disks.

\subsection{G\"ultekin sample}

\begin{figure*}
\centerline{
\includegraphics[angle=0.,width=0.5\hsize]{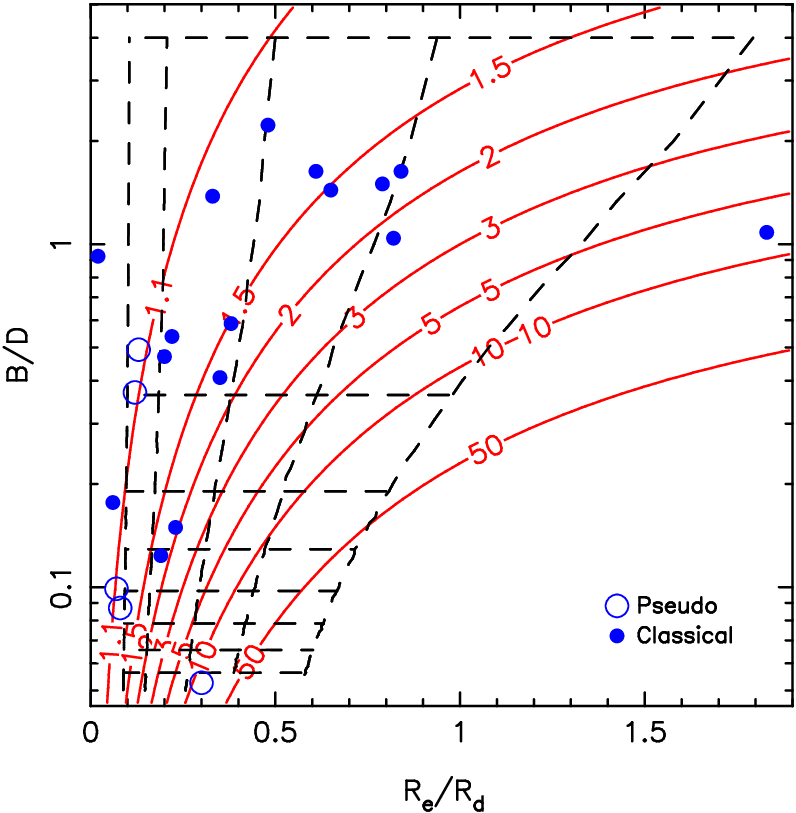}
\includegraphics[angle=0.,width=0.5\hsize]{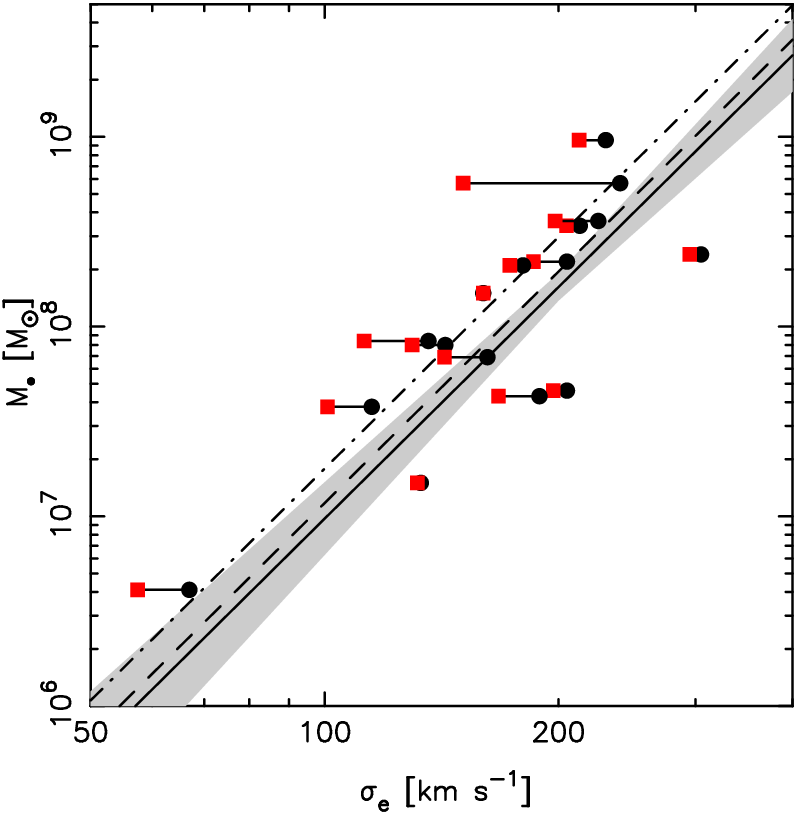}
}
\caption{The sample of \citet{gultekin+09} with bulge$+$disk fits as
  described in the text.  Left panel: Distribution of the sample in
  the \bd-\rerd\ plane.  Filled symbols show $n>2$ (classical) bulges
  while open symbols show $n<2$ (pseudo) bulges.  Contours of \dMbh\
  assume $\beta = 4.24$, as before.  Right panel: The \Msig\ relation
  for the 16 classical bulges.  The (black) circles with larger values
  mark the observed \sig{e}\ while the connected (red) squares show
  \sig{e,0}\ (i.e., the decompressed values) from the fit of Eqn.
  \ref{eqn:fit}.  The solid line shows our re-fit for the \Msig\
  relation of elliptical galaxies ($\alpha = 8.21$, $\beta = 4.06$)
  while the dashed and dot-dashed lines show the observed ($\alpha =
  8.29 \pm 0.09$) and decompressed ($\alpha = 8.47 \pm 0.11$)
  classical bulges fitted by a relation with $\beta = 4.06$.  The
  shaded region shows the one sigma uncertainty on the \Msig\ relation
  of ellipticals.
  \label{fig:gultekin}}
\end{figure*}

\citet{gultekin+09} presented a sample of 49 galaxies with \Mbh\
measurements, to which they fitted the \Msig\ relation.  Many of these
galaxies have bulge$+$disk decompositions in the literature
\citep{fis_dro_08, fisher_drory10, fisher_drory11}.  The photometric
decompositions we use here were taken from \citet{fis_dro_08},
\citet{fisher_drory10} and \citet{fisher_drory11}.  For the
unpublished decompositions the Appendix provides a description of how
they were performed.  The left panel of Figure \ref{fig:gultekin}
presents these photometric decompositions with contours of \dMbh\ from
Eqn. \ref{eqn:fit} overlaid.

\citet{gultekin+09} found $\alpha = 8.23$ for ellipticals and $\alpha
= 8.17$ for classical bulges, but their definition of classical bulges
includes the elliptical galaxies.  We therefore refit the relation to
ellipticals and classical bulges separately using the code {\sc
  mpfitexy}\footnote{http://purl.org/mike/mpfitexy} which implements
the MPFIT algorithm \citep{markwardt09}.  We first fit the \Msig\
relation of the elliptical galaxies in the \citet{gultekin+09} sample:
IC 1459, M32, M60, M84, M87, NGC 821, NGC 1399 (both measurements),
NGC 2778, NGC 3377, NGC 3379, NGC 3607, NGC 3608, NGC 4261, NGC 4291,
NGC 4459, NGC 4473, NGC 4486A, NGC 4697, NGC 5077, NGC 5576, NGC 5845,
NGC 6251, NGC 7052, A1836 and A3565.  We obtain $\beta = 4.06 \pm
0.40$, with zero-point $\alpha = 8.21 \pm 0.07$ and an intrinsic
scatter of 0.30.  This measurement is in excellent agreement with the
\Msig\ relation of elliptical galaxies obtained by \citet{gultekin+09}
using a different fitting method.  This \Msig\ relation is shown by
the solid line in the right panel of Figure \ref{fig:gultekin}.  We
then selected the classical bulges to be those having $n>2$, leaving
us with 16 galaxies: NGC 224, NGC 1023, NGC 2787, NGC 3031, NGC 3115,
NGC 3227, NGC 3245 NGC 3585, NGC 3998, NGC 4026, NGC 4258, NGC 4342,
NGC 4564, NGC 4594, NGC 4596 and NGC 7457.  For this sample we
measure, from the photometric decompositions, $\left<{\cal{F}}\right>
= 1.12 \pm 0.03$, comparable to the values predicted in Table
\ref{tab:ffactors}.  Based on this value and fixing $\beta = 4.06$, we
expect $\delta\alpha = -0.20$ if \Mbh\ had not changed as the disks
grew.  Fitting the \Msig\ relation for classical bulges while holding
$\beta$ fixed, we obtain $\alpha = 8.29 \pm 0.09$, which is plotted as
the dashed line in Figure \ref{fig:gultekin}.  The offset from the
elliptical relation is only 0.08, within the one sigma uncertainty and
significantly smaller than expected from the photometric decomposition
if no SMBH growth had occurred.  The zero-point predicted by the
photometric decompositions is $\sim 2$ sigma away from the one found.
Thus we find no evidence of a significant offset between elliptical
galaxies and observed classical bulges.

In order to demonstrate that bulge compression should have produced an
offset that is measureable, we also fitted the \Msig\ relation of the
same bulges decompressed using Eqn. \ref{eqn:fit}, again fixing $\beta
= 4.06$ and using as uncertainties on \sig{e,0} and \Mbh\ the values
for the observed bulges.  We obtain $\alpha = 8.47 \pm 0.11$.  The
offset from the relation for ellipticals is now $+0.26$, which is two
sigma different.  This fit is shown in the right panel of Figure
\ref{fig:gultekin} as the dot-dashed line.

\subsection{Beifiori sample}

\begin{figure*}
\centerline{
\includegraphics[angle=0.,width=0.5\hsize]{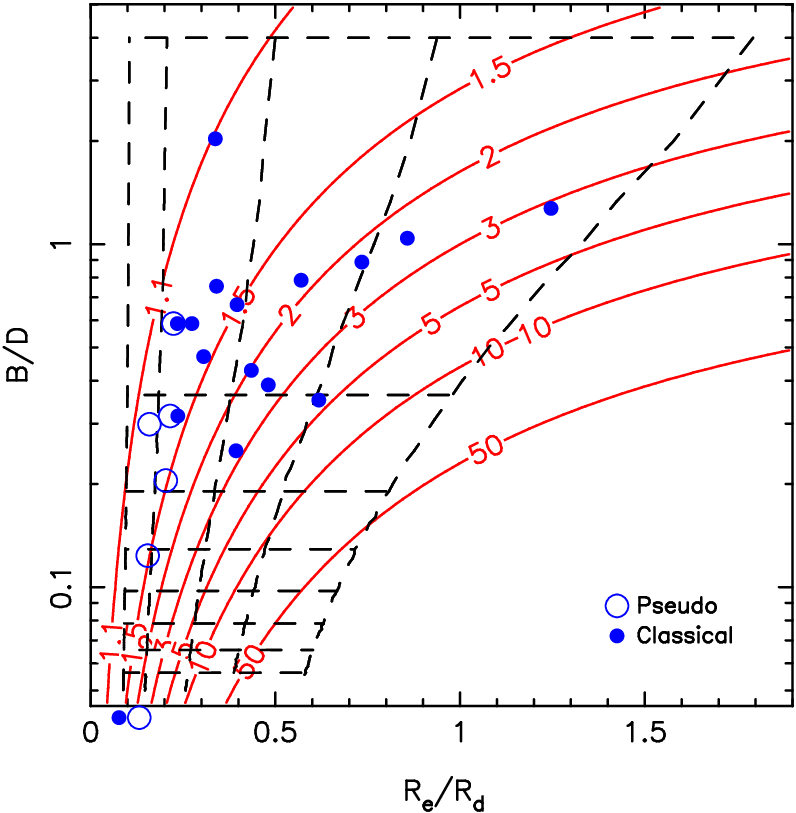}
\includegraphics[angle=0.,width=0.5\hsize]{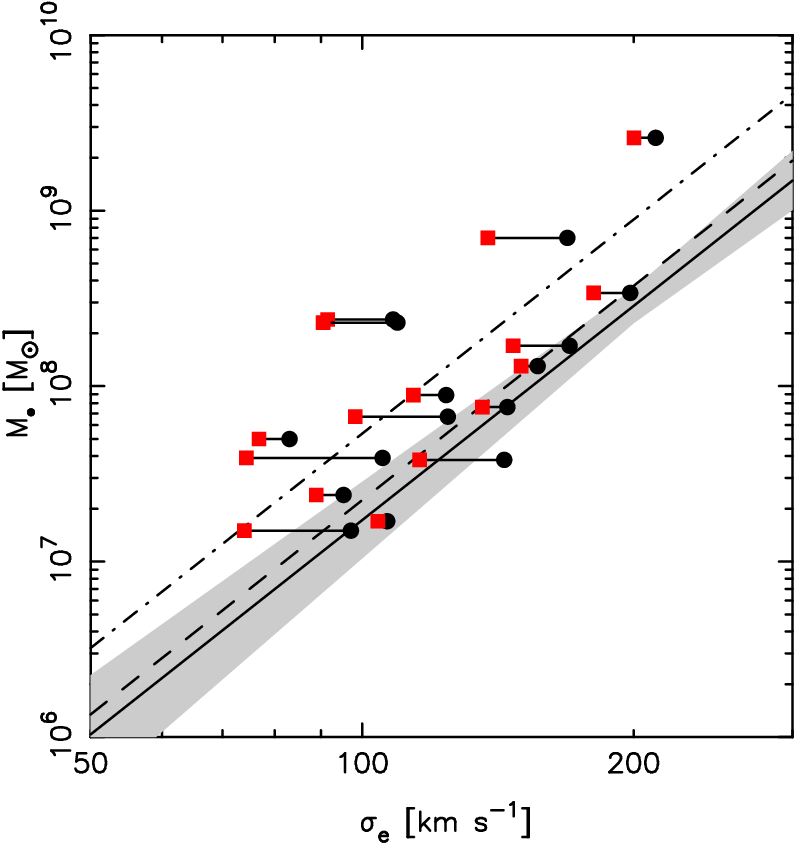}
}
\caption{ The sample of galaxies with upper limits on \Mbh\ from
  \citet{beifiori+09}.  Left panel: Distribution of the sample in the
  \bd-\rerd\ plane.  Filled symbols show $n>2$ (classical) bulges
  while open symbols show $n<2$ (pseudo) bulges.  Contours of \dMbh\
  assume $\beta = 4.24$ \citep{gultekin+09}.  Right panel: The \Msig\
  relation of the 16 classical bulges.  All black hole masses are
  upper limits only.  The (black) circles with larger values mark the
  observed \sig{e}\ while the connected (red) squares show \sig{e,0}\
  (i.e., the decompressed values) from the fit of Eqn. \ref{eqn:fit}.
  The various lines show our fits of the \Msig\ relation to different
  samples with the slope of the relation held fixed to that for
  elliptical galaxies in the \citet{gultekin+09} sample ($\beta =
  4.06$).  The solid line shows the fit to the ellipticals ($\alpha =
  8.46 \pm 0.10$), the dashed line fits the observed classical bulges
  ($\alpha = 8.57 \pm 0.10$) and the dot-dashed line the decompressed
  classical bulges ($\alpha = 8.95 \pm 0.11$).  The shaded region
  shows the one sigma uncertainty on the \Msig\ relation of
  ellipticals.
\label{fig:beifiori}}
\end{figure*}

As a further demonstration of the absence of an offset in the \Msig\
relation between ellipticals and classical bulges we consider also the
independent sample of \citet{beifiori+09}.  \citet{beifiori+09}
obtained upper limits on the masses of SMBHs in over 100 galaxies.
They showed that their relation is parallel to the usual \Msig\
relation, with $\beta = 4.12 \pm 0.38$.  For a number of these
galaxies, \citet{beifiori+12} provide bulge$+$disk decompositions; the
resulting sample has 22 disk galaxies.  Of these, 16 galaxies have
$n>2$ which we select as classical bulges: NGC 2911, NGC 2964, NGC
3627, NGC 3675, NGC 3992, NGC 4203, NGC 4245, NGC 4314, NGC 4429, NGC
4450, NGC 4477, NGC 4548, NGC 4579, NGC 4698, NGC 5005 and NGC 5252.
The left panel of Figure \ref{fig:beifiori} plots the distribution of
these bulges in the \bd-\rerd\ plane.  From their photometric
decompositions we obtain $\left<{\cal{F}}\right> = 1.16 \pm 0.03$.

Because this sample only has upper limits on \Mbh, not actual
measurements, we fit the \Msig\ relation keeping the slope of the
relation fixed to that obtained for the elliptical galaxies from the
\citet{gultekin+09} sample, \ie\ $\beta = 4.06$.  Although the
\citet{beifiori+09} sample of upper limits cannot give the absolute
zero-point of the relation, we are interested in relative offsets, for
which it is well suited.  As an estimate for the error on \Mbh\ we use
half the difference between the two upper limits given by
\citet{beifiori+09}, which are based on assuming two different
inclinations for the nuclear disk surrounding the SMBH.  Using a
constant error of $10^3 \Msun$ instead yields results that are
virtually indistinguishable. We fit the zero-points for ellipticals
($\alpha = 8.46 \pm 0.10$), observed classical bulges ($\alpha = 8.57
\pm 0.10$) and decompressed classical bulges ($\alpha = 8.95 \pm
0.11$).  These results are shown in the right panel of Figure
\ref{fig:beifiori}.  The offset between the ellipticals and the
observed classical bulges is 0.11, which is again less than one sigma.
In comparison, the offset between ellipticals and decompressed
classical bulges, shown in the right panel of Figure
\ref{fig:gultekin}, is $+0.49$ (the photometric prediction being
$\delta \alpha = 0.26$), which is different at more than three sigma.
Hence, for this sample the observed offset and the offset predicted if
no SMBH growth occurs differ by $> 3$ sigma.

Thus both the \citet{gultekin+09} sample, with full \Mbh\
measurements, and the \citet{beifiori+09} sample with \Mbh\ upper
limits only, show no evidence for an offset between the \Msig\
relation of ellipticals and of classical bulges, even though the disks
should have compressed the bulges to a measureable extent.  We
therefore conclude that SMBHs in classical bulges have been growing
along with disks.

\subsection{Galaxies with $\dMbh > 3$}

We estimated above that the scatter in $\sig{e}/\sig{e,0}$ is $\la
30\%$.  This implies that galaxies in which ${\cal{F}} > 1.3$, (\ie\
$\dMbh > 2.9$ for $\beta = 4$) should be dominated by compression.
Table \ref{tab:dMbhs} lists the five galaxies for which the
photometric properties imply $\dMbh \geq 3$; these are the galaxies
for which the impact of compression is the largest, and are therefore
ideally suited to test whether or not disk (re-)assembly is associated
with growth of the SMBH.  Of the five galaxies, only one, NGC 4594
(the Sombrero galaxy), which happens to have the largest \dMbh, has a
proper \Mbh\ mass measurement.  Two of the other galaxies have \Mbh\
upper limits from \citet{beifiori+09}.  The remaining two galaxies
have no \Mbh\ measurements that we are aware of.  We recommend
measurements of \Mbh\ in these galaxies in order to further constrain
the ability of SMBHs to grow along with disks.

\begin{table}[!ht]
\begin{centering}
\begin{tabular}{ccl} \\ \hline
\multicolumn{1}{c}{Galaxy} &
\multicolumn{1}{c}{\dMbh} &
\multicolumn{1}{l}{Reference} \\ \hline
 NGC 4594 &  6.6 & \citet{fisher_drory11} \\
 NGC 3675 &  4.1 & \citet{beifiori+12}    \\
 NGC  438 &  3.1 & \citet{graham_03}      \\
 NGC 3627 &  3.0 & \citet{beifiori+12}    \\
 NGC 3140 &  3.0 & \citet{graham_03}      \\ \hline
\end{tabular}
\caption{Galaxies for which the photometric data predicts $\dMbh \geq
  3$ assuming the \Msig\ relation of ellipticals ($(\alpha,\beta) =
  (8.21,4.06)$).  The column labelled `Reference' lists the source for
  the bulge$+$disk decomposition.
  \label{tab:dMbhs}}
\end{centering} 
\end{table}

\begin{figure}
\centerline{
\includegraphics[angle=-90.,width=\hsize]{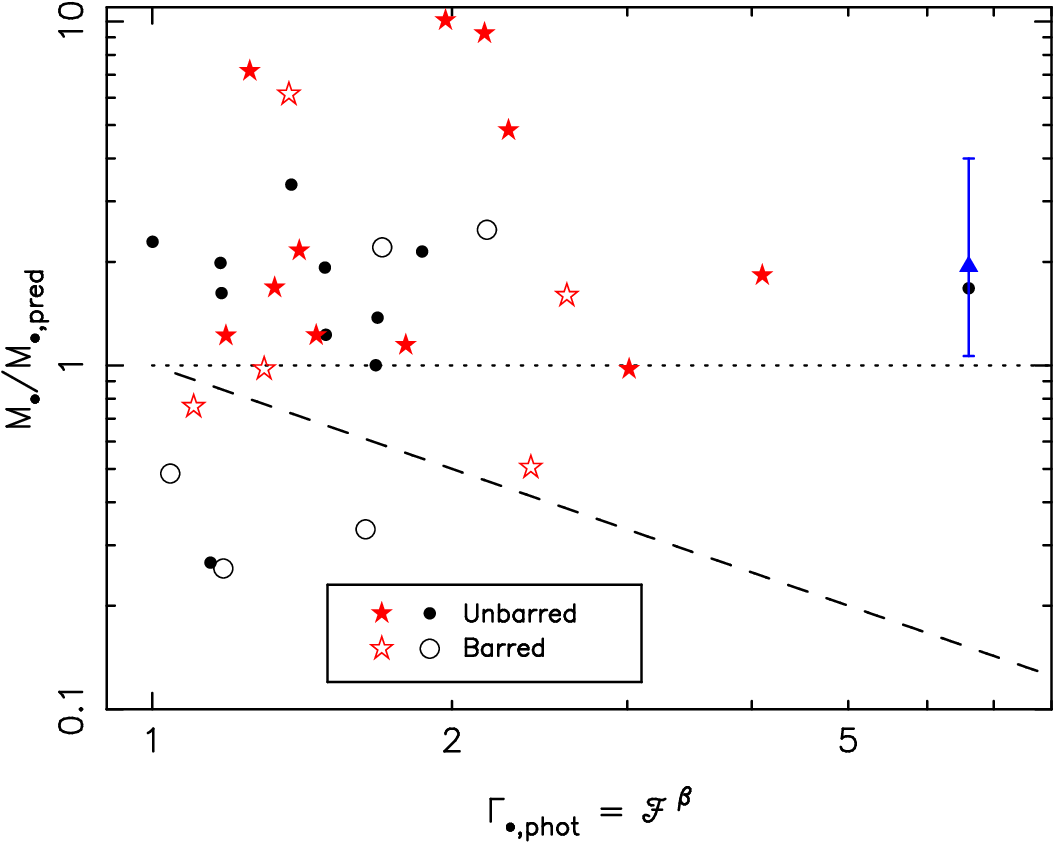}
}
\caption{The residuals from (our fit for) the \Msig\ relation for
  ellipticals ($\alpha=8.21$ for the \citet{gultekin+09} sample and
  $\alpha = 8.46$ for the \citet{beifiori+09} sample, with
  $\beta=4.06$ for both) versus \dMbh\ predicted by the photometric
  decompositions. Barred and unbarred galaxies are shown as open and
  filled symbols, respectively.  The (black) circles are SMBHs from
  \citet{gultekin+09} while \dMbh\ is computed using the bulge$+$disk
  decomposition of \citet{fisher_drory10} (see text for details).  The
  (red) stars are SMBH upper limits from the sample of
  \citet{beifiori+09}, with decompositions from \citet{beifiori+12}.
  The (blue) triangle with error bars shows NGC 4594 (the Sombrero
  galaxy) with improved \Mbh\ measurement taken from
  \citet{jardel+11}.  The dashed line shows $\Mbh/{\rm
    M}_{\bullet,{\rm pred}} = \dMbh^{-1}$ while the dotted line shows
  $\Mbh/{\rm M}_{\bullet,{\rm pred}} = 1$. 
\label{fig:residuals}}
\end{figure}

Figure \ref{fig:residuals} plots the ratio of the observed \Mbh\ to
that predicted for its value of \sig{e}\ by the \Msig\ relation of
ellipticals versus \dMbh$_\mathrm{,phot} \equiv {\cal F}^\beta$, the
value of \dMbh\ predicted by the photometric decompositions.  The
dashed line showing $\Mbh/{\rm M}_{\bullet,{\rm pred}} =
\dMbh_\mathrm{,phot}^{-1}$ represents the location of SMBHs that form
on the \Msig\ relation and do not grow as the disk regrows.  The
dotted line instead shows the case $\Mbh/{\rm M}_{\bullet,{\rm pred}}
= 1$, corresponding to SMBHs that always stay on the \Msig\ relation
as the disk regrows.  Most galaxies are above or near the dotted line,
and this is especially true at $\dMbh_\mathrm{,phot} > 3$, regardless
of whether the \citet{gultekin+09} or the \citet{beifiori+09} sample
is considered.  For the galaxy with the largest predicted
\dMbh$_\mathrm{,phot}$, NGC 4594, we also plot the improved \Mbh\
measurement of \citet{jardel+11} together with its uncertainty.  NGC
4594 provides the greatest leverage in distinguishing how SMBHs and
disks co-evolve; Figure \ref{fig:residuals} shows clearly that its
SMBH continued to grow while its disk was forming.  Galaxies would
have followed the dashed line in Figure \ref{fig:residuals} if the
\Mmb\ had been the more fundamental scaling relation rather than the
\Msig\ relation, as assumed here.

There is a hint that barred galaxies are more frequently found near or
below the dashed line in Figure \ref{fig:residuals}, although this is
not true of all barred galaxies.  However the data do not reach
\dMbh$_\mathrm{,phot}$\ values large enough to determine whether there
is a real difference between barred and unbarred galaxies.

%%%%%%%%%%%%%%%%%%%%%%%%%%%%%%%%%%%%%%%%%%%%%%%%%%%%%%%%%%%%%%%%%%%%%

\section{Discussion and Conclusions}
\label{sec:conclusions}

Observations find that the peak of the integrated AGN activity is at
$z \simeq 2$ \citep{wolf+03}.  The majority of bright quasars are in
elliptical galaxies \citep{kukula+01, dunlop+03, kauffmann+03} but
intermediate brightness Seyfert AGN, which represent a significant
fraction of the total AGN number density at $z = 1.5-3$
\citep{ueda+03}, are preferentially in disk galaxies
\citep{schawinski+11}.  \cite{schawinski+11} estimate that $23-40\%$
of SMBH growth in these AGN occur during a slow, secular mode of the
type envisaged here.  For the samples of disk galaxies with classical
bulges that we explored here we estimate a mean \Mbh\ growth by $\sim
50\%-65\%$.  If instead we consider the samples with \Mbh\
measurements (both the direct and upper limits only) we find growth
factors $\sim 60\%-80\%$ from their photometric decompositions.  The
growth factor may be somewhat larger still if disks are steeper than
exponential at their centers.  Nonetheless, our estimated growth
factor spans a range that is broadly in agreement with observational
estimates.

We failed to find a significant difference between the \Msig\ relation
of ellipticals and of classical bulges.  With currently available
samples this result is statistically significant only at about
two-three sigma.  Besides increasing the sample size, the best future
prospects for improving the significance of this result is if more
galaxies with photometrically predicted large values of \dMbh\ were to
have their \Mbh\ measured.  We have provided a list of 5 galaxies
(Table \ref{tab:dMbhs}) with $\dMbh \geq 3$; of these 4 have no
directly measured \Mbh.  Galaxies with such large predicted growth
factors offer excellent probes of the co-evolution of SMBHs and disks.
Moreover Eqn. \ref{eqn:fit} makes it easy to trawl through photometric
catalogs to search for further examples of galaxies with large
predicted growth factors.

Since this paper was first submitted there have been several updates
to the \citet{gultekin+09} sample used in this work. We explored the
impact of these via the sample compiled in \citet{mcconnell_ma12}.
The main changes for elliptical galaxies were updates of some SMBH
masses and \sig{e}'s and the addition of SMBH measurements in several
brightest cluster galaxies (BCGs).  Because BCGs evolve differently,
we exclude these new galaxies from our sample and use the same sample
as listed above under the \citet{gultekin+09} sample, updating to the
new \Mbh\ and \sig{e}\ values (dropping NGC 2778 which does not have a
significant SMBH detection in recent measurements).  For this sample
we obtain $(\alpha,\beta) = (8.37\pm0.07,4.39\pm0.42)$.  The
\citet{mcconnell_ma12} sample includes a number of new SMBH
measurements in disk galaxies; at present we cannot determine whether
any of these galaxies host classical bulges.  In any case, several of
these are low mass galaxies and are likely to host pseudo bulges, so
we continue to fit to the same classical bulge sample from
\citet{gultekin+09} as before, now fixing the slope to $\beta=4.39$.
We obtain an intercept $\alpha = 8.32 \pm 0.09$, which is
statistically indistinguishable from the value for ellipticals.
Instead the value for the decompressed bulges is $\alpha = 8.51 \pm
0.11$, one sigma different from the value for ellipticals.  We
conclude that the latest measurements continue to show no evidence of
an offset between ellipticals and classical bulges.

Assuming our result continues to hold with increased sample size, the
consequence of our finding is that SMBHs grow along with disks.  The
main parameter regulating their growth is then the potential within
which they reside, which is largely set by the bulge.  This means that
SMBH growth is self-regulated \citep[e.g.][]{treister+11}: SMBHs can
grow until their feedback unbinds any gas otherwise destined to
accrete onto them.  This picture accounts also for the absence of
correlations with properites of the dark matter halo or of the disk
\citep{kormendy_bender_11, kormendy_bender_cornell_11}.

Disk mass growth leads to an evolution of \sig{e}\ that is
non-hierarchical, thereby adding nothing to the mass of a classical
bulge.  Another consequence of the absence of an offset in the \Msig\
relation of classical bulges therefore is that the bulge mass, which
does not change as \db\ increases, is not the main parameter
determining \Mbh.  Thus the \Mmb\ relation cannot be as fundamental as
the \Msig\ relation.  One interpretation of SMBH scaling relations
views them as reflecting only a central-limit-theorem non-causal
evolution produced by repeated galaxy merging \citep{peng07,
  jahnke_maccio11}.  In this picture the main correlation is between
\Mbh\ and \Mbul, both of which grow during mergers.  \citet{peng07}
even predicted that bulge-dominated galaxies will have tighter scaling
relations than disk-dominated ones.  The lack of an offset between
ellipticals and classical bulges is contrary to this scenario: some
form of regulation between SMBHs and bulges is required.

\subsection{The Milky Way Galaxy}

Whether the Milky Way hosts a classical or pseudo bulge remains
unclear.  While its bulge stars are mostly old, metal-rich and
$\alpha$-enhanced, favoring fast formation during mergers
\citep{mcwilliam_rich_94, zoccali+04, zoccali+06, zoccali+08,
  lec_etal_07, fulbright+07}, kinematics and morphology favor its
formation via the central bar (\citet{fux_97, fux_99, jshen+10} but
see also \citet{saha+12}).  Assuming it is a classical bulge, the
green star in the right panel of Figure \ref{fig:deltaMs} represents
the Milky Way based on a bulge$+$disk decomposition of the density
model of \citet{bis_ger_02} ($\db = 8.3$, $\rdre = 3.2$); this implies
$\dMbh \simeq 3.7$.  If currently $\Mbh = 4.1 \times 10^6~ \Msun$
\citep{ghez+08, gillessen+09}, the original SMBH would have had $\Mbh
\sim 1.1 \times 10^6~ \Msun$ if it formed on the \Msig\ relation.

\subsection{Summary}

Our main results can be summarized as follows:

\begin{enumerate}

\item When a disk forms and grows around a pre-existing bulge, it
  gravitationally compresses the bulge, causing its effective velocity
  dispersion, \sig{e}, to increase.  We have provided a fitting
  formula, Eqn. \ref{eqn:fit}, for the change in \sig{e}\ for given
  bulge-to-disk mass and size ratios.

\item Using the SDSS data of \citet{gadotti09} and
  \citet{gadotti_kauffmann_09}, we find evidence that classical bulges
  have been compressed as disks reformed around them.  The photometric
  samples predict that bulges should experience a mean increase in
  \sig{e}\ by $\sim 10\%$.  While small, the steepness of the \Msig\
  relation requires SMBHs to grow, on average, by $\sim 50\%$ and
  extends to $> 200\%$.

\item The weak correlations between \db\ and \sig{8}\ and between
  \rdre\ and \sig{8}\ ensure that the main effect of bulge compression
  on the \Msig\ relation, if \Mbh\ remains unchanged as the disk
  regrows, is an offset to a smaller zero-point at fixed slope.  The
  predicted offset between ellipticals and classical bulges is
  measureable with available samples of \Mbh.

\item We do not find an offset between the \Msig\ relations of
  ellipticals and of classical bulges in either the sample of
  \citet{gultekin+09} or that of \citet{beifiori+09}.  Using available
  photometric decompositions of the galaxies, we show that an offset
  should have been found if \Mbh\ had not changed since the bulges
  formed.  {\em Thus SMBHs must have grown along with disks.}

\item We estimate that SMBHs had to have grown by $\sim 50\%-80\%$ in
  order to remain on the \Msig\ relation.  Such significant SMBH growth
  is in agreement with recent observations that find that at $1.5 \le
  z \le 3$ SMBHs in disk galaxies grow by $\sim 23\%-40\%$.

\item We have provided a list of 5 galaxies (Table \ref{tab:dMbhs})
  for which the SMBH is predicted to have needed to grow by a factor
  greater than three to remain on the \Msig\ relation.  SMBHs with
  such large growth factors provide strong constraints on the
  mechanisms regulating the \Msig\ relation and we strongly encourage
  measurement of their black hole masses.

\end{enumerate}

%%%%%%%%%%%%%%%%%%%%%%%%%%%%%%%%%%%%%%%%%%%%%%%%%%%%%%%%%%%%%%%%%%%%%

%%%%%%%%%%%%%%%
% Acknowledgments
%%%%%%%%%%%%%%%

\acknowledgments We are very grateful to Alessandra Beifiori and David
Fisher for providing us with data prior to publication.  Additional
data were kindly provided by Dimitri Gadotti.  We thank Alessandra
Beifiori and Monica Valluri for useful discussions.  We especially
thank Markus Hartmann for fitting the \Msig\ relation for various
samples in this paper.  We thank the anonymous referee for a useful
report that helped to improve this paper.  S.K. is funded by the
Center for Cosmology and Astro-Particle Physics (CCAPP) at The Ohio
State University. F.B.  acknowledges financial support from the Lady
Davis Fellowship Trust.  The simulations in this paper were carried
out at Albert, the supercomputer at the University of Malta and on the
COSMOS Consortium supercomputer within the DIRAC Facility jointly
funded by STFC, the Large Facilities Capital Fund of BIS.  This
research was also supported by an allocation of computing time from
the Ohio Supercomputer Center (http://www.osc.edu).

%%%%%%%%%%%%%%%%%%%%%%%%%%%%%%%%%%%%%%%%%%%%%%%%%%%%%%%%%%%%%%%%%%%%%

\appendix
\section{Unpublished photometric fits}

We make use of unpublished decompositions of disk galaxies in the
sample of \citet{gultekin+09} kindly provided to us by David Fisher.
Most of these decompositions have been published \citep{fis_dro_08,
fisher_drory10, fisher_drory11}.  However a few remain unpublished and
we provide here a description of the analysis method by which David
Fisher derived these decompositions.  

The decompositions use archival {\it HST} and ground-based data.  When
possible, near infrared data are used as they are less sensitive to
the obscuring effects of dust.  \citet{fis_dro_08} show that for
relative quantities, such as $B/T$, there is little difference from
$V$-band to $H$-band; data are therefore restricted to be $V$-band or
redder. For each galaxy the surface brightness profile is determined
through ellipse fitting of both {\it HST} and ground-based data,
thereby simultaneously constraining both the small scale structure at
the center of the galaxy and the shape of the outer disk profile.
Interfering objects, such as foreground stars and background galaxies,
are masked via automatic source identification methods and manually
removed. For ground-based images the sky is removed by subtracting a
surface, fitted to regions of images that do not contain galaxy light.
The radial sizes of the ellipses are optimized to maintain a roughly
constant signal-to-noise ratio across the profile, and zero-point
shifts of the ground-based image to match the {\it HST} data ensured
continuity. The bulge$+$disk decompositions are then determined by
fitting a S\'ersic bulge plus outer exponential disk to the major axis
surface brightness profile.

%\bibliographystyle{aj.bst}
%\bibliography{allrefs}
\bibliographystyle{aj}
\bibliography{debattista}

%%%%%%%%%%%%%%%
% End of Document
%%%%%%%%%%%%%%%

\end{document}